# Evaluating Organizational Effectiveness: A New Strategy to Leverage Multisite Randomized Trials for Valid Assessment

Guanglei Hong[1], Jonah Deutsch[2*], Peter Kress[2*], Jose Eos Trinidad[3*], Zhengyan Xu[4*]

[1] University of Chicago; [2] Mathematica; [3] University of California-Berkeley; [4] University of Pennsylvania

* Names of the co-authors are in alphabetical order.

**Abstract:**

Determining which organizations are more effective in implementing an intervention program is essential for theoretically and empirically characterizing exemplary practice and for intervening to enhance the capacity of ineffective ones. However, sites differ in their local ecological conditions including client composition, availability of alternative programs, and broader community context. Applying the causal inference framework, this study proposes a formal mathematical definition of an organization's local relative effectiveness attributable solely to malleable organizational practice. Capitalizing on multisite randomized trials, the identification leverages observed control group outcomes that capture some of the confounding impacts of otherwise unmeasured contextual variation. We propose a two-step mixed-effects modeling (2SME) procedure that adjusts for pre-existing between-site variation. Monte Carlo simulations demonstrate its superior performance compared to conventional methods. We apply the new strategy to an evaluation of Job Corps centers nationwide serving disadvantaged youths.

Many social services are delivered by local organizations across multiple sites. Determining which organizations best achieve program goals is crucial for building theory on malleable organizational practice that drives effectiveness. Such theoretical knowledge will help policymakers identify and support organizations in need of capacity building (Besharov et al., 2016; Bloom, 2005; Century & Cassata, 1996; Patton, 2001; Russ-Eft et al., 2024; Walton et al., 2019; Weiss et al., 2014).

Familiar concepts such as the site-specific intent-to-treat (ITT) effects[1] are suitable parameters for evaluating an intervention program in contrast with the local "business as usual" in serving a local population. These causal effects are useful for informing intervention program developers and funding agencies in deciding whether to invest in the program and where to operate. However, they are not suitable for evaluating the relative effectiveness of organizations independent of variation in the populations they serve and other local conditions.

Prior work has demonstrated that using these familiar site-specific ITT effects fails to isolate organizational effectiveness when the latter is the key parameter of interest. This is because the site-specific ITT effect reflects not only how well an organization serves individuals in the treated condition but also variations in experiences and outcomes of untreated individuals across sites (Cordray & Pion, 1993; Lipsey & Cordray, 2000; Raudenbush & Willms, 1995; Rogers, 2016). As Weiss and colleagues (2014) noted, local ecological conditions including client composition, available alternative programs, and community context represent pre-existing heterogeneity across sites that may distort the assessment of organizational effectiveness. Several hypothetical examples may clarify the potential significance of such confounding factors.

[1] We limit attention to cases in which the local organization does not serve the untreated group. For ease of exposition, we discuss the site-specific ITT effect here as a familiar parameter, though without loss of generality the discussion could be broadened to also include the site-specific treatment-on-the-treated effect, local average treatment effect, and others.

– At a site in which eligible individuals experience greater disadvantages and require investments in extra support when compared with those at most other sites, an organization that manages to effectively overcome a constellation of barriers may display an unimpressive ITT effect in comparison with peer organizations at sites where extra support for clients is not needed.

– At a site where high-quality *alternative* programs are available under the control condition and abundant *opportunities* exist for attaining desired outcomes, a high-performing organization serving experimental individuals may show a negligible ITT effect. In contrast, when few good alternatives are available, a mediocre organization serving the experimental individuals may show an impressive ITT effect simply due to the inadequacy of the "business as usual" at the site.

Built upon past research, this study proposes conceptual and methodological approaches to isolate the relative effectiveness of a local organization in comparison with peer organizations operating under similar local ecological conditions. Applying the causal inference framework, we provide a formal mathematical definition, highlighting the need to attend to contextual variation in attaining a valid evaluation, which is essential for advancing theoretical understanding about organizational improvement.

Leveraging multisite randomized trials (Westine, 2016), we clarify identification conditions and develop a new analytic strategy. Our key insight is that, in a multisite randomized trial, control group outcomes at each site summarize between-site differences in a host of unmeasured confounding factors such as prior vulnerability shaped by individuals' earlier life experiences and community conditions including local unemployment rates. We view an individual's potential outcome associated under the control condition as reflecting

“predisposition-by-environment interaction” in the absence of intervention. It is a proxy for joint influences of measured and unmeasured individual characteristics, the accessibility and quality of alternative programs, and subsequent opportunities or constraints in the local community context. When the treatment assignment is randomized at a site, the empirical distribution of the observed outcomes in the control group captures some of the otherwise unmeasured information about the influences of these local ecological conditions in the absence of the program.

To implement, this paper lays out a two-step mixed-effects (2SME) modeling procedure that adjusts for between-site differences in not only observed client composition at the baseline but also the observed control group outcomes. In the absence of other unmeasured confounders, this strategy isolates the program impact on the outcomes of the target population of clients solely attributable to the relative effectiveness of an organization. Monte Carlo simulations demonstrate the superior performance of this new strategy when compared to existing methods that rely entirely on the site-specific ITT effect estimates or adjust only for between-site differences in observed baseline client composition. We apply the new strategy to a reanalysis of data from the National Job Corps Study.

## Relevant Past Research

Decades of organizational studies have suggested that internal contributing factors to the effectiveness of an organization may involve organizational leadership (Bass & Avolio, 1994; Yuki, 2008), professional knowledge and skills of staff (Shullman, 1986, 2011), their shared beliefs and collective sensemaking (Coburn, 2001; Spillane et al., 2002; Weick, 1987), relational trust (Bryk & Schneider, 2002), competence at organizational learning (Levitt & March, 1988), and implementation as mutual adaptation (Leonard-Barton, 1988; McLaughlin, 1976). Credible evaluation evidence is in demand for determining what organizational practice to stop, continue, or scale. However, to empirically examine theoretical arguments regarding best organizational practice, the first-order question is how to rigorously define and quantify organizational effectiveness (Benjamin, 2012; Cameron & Whetten, 1983; Lee & Nowell, 2015; Sharma & Singh, 2019; Steers, 1975).

Conventional performance management typically ranks organizations based on their performance records rather than true impact. These comparisons are confounded by pre-existing contextual differences between sites and may incentivize "gaming" practices such as "cream skimming," where organizations select individuals who would likely succeed without the program (Anderson et al., 1993; Barnow, 1992; Heckman et al., 2002; Heckman et al., 2011). Although when the service is oversubscribed at a site, lottery-type randomization of treatment assignment eliminates within-site selection bias, confounding between sites persists.

To address conceptual confusion surrounding value-added effects of school organizations, Raudenbush and Willms (1995) contrast two different parameters suited for different purposes. The "Type A effect," relevant to parents choosing schools, reflects both school practices and contextual influences beyond the immediate control of school staff. In contrast, the "Type B effect" isolates the value added by school practices alone and is appropriate for accountability. As they define it, the Type B effect is "the difference between a child's performance in a particular school and the performance that would have been expected if that child had attended a school with identical context but with practice of "average" effectiveness" (p.310).

According to Raudenbush and Willms, identifying the Type B effect would require assigning students at random to schools and, additionally, assigning schools at random to different levels of organizational effectiveness. As the latter rarely happens, they conclude that analytic strategies are lacking for identifying the Type B effect. Indeed, various value-added measures typically fail to adjust for unmeasured pre-existing differences across local contexts with notable consequences (Barnow & Heinrich, 2010; Raudenbush, 2004; Reardon & Raudenbush, 2009; Rubin et al., 2004; Schochet & Fortson, 2014).

Multisite randomized trials are well-suited for identifying and consistently estimating site-specific ITT effects and their distribution across sites (Weiss et al., 2014). Past research has additionally sought to explain between-site variation in ITT effects by examining site-level characteristics as moderators (Bloom, 2005; Bloom et al., 2003; Bloom et al., 2017; Cox & Kelcey, 2023; Morris et al., 2018; Olsen et al., 2024; Raudenbush & Bloom, 2015; Seltzer, 1994). These moderation studies test hypotheses about who may benefit from the intervention program and under what conditions. However, using the site-specific ITT effect as a proxy for the relative effectiveness of a local organization conflates the Type A and Type B effects. A related line of research has centered on generalizing or transporting findings about the program benefit from a randomized trial at a local site to a broader population (Dahabreh et al., 2019, 2020, 2024; Degtiar & Rose, 2023; Tipton & Olsen, 2018). However, the literature on transportability assumes that treatment effects for subpopulations do not depend on other local conditions. Flores and Mitnik (2013) challenge this strong assumption, highlighting the salience of posttreatment local economic conditions in moderating the treatment effect, though their approach requires time-series measures of local conditions. Extending the idea of transportability, Lu and colleagues (2023) propose strategies to equate the distribution of observed baseline individual-level characteristics at each site with that of an arbitrary target population, partially parceling out some between-site differences in populations served. Still, residual between-site variation in ITT effects remains confounded by access to alternative programs, community context, and unobserved compositional differences. As Lu and colleagues (2023) note, "A principled framework for cross-site treatment effect comparisons is needed to disentangle these sources of variation" (p.22). This study aims to further develop such a principled framework.

To identify the Type B effect — that is, the relative effectiveness of a local organization attributable to its practices given its local ecological conditions, we capitalize on the within-site treatment randomization in a multisite randomized trial. Our identification strategy creatively leverages control group outcomes at each site that capture some of the otherwise unmeasured site-level differences in the local ecological conditions.

## Job Corps Evaluations

To illustrate the value of our analytic innovations, we apply them to an evaluation of the relative effectiveness of individual Job Corps centers. Job Corps is a major federally funded program in the U.S. that provides job training and other services to disadvantaged youths between ages 16-24 who face multiple barriers in completing school and entering the workforce (McConnell & Glazerman, 2001). An effective Job Corps center is expected to produce significant positive changes in the lives of eligible youth (Burghardt & Schochet, 2001). In the mid-1990s, the National Job Corps Study (NJCS) included all the one hundred or so Job Corps centers nationwide in a multisite randomized trial. At each site, eligible applicants were assigned at random to either enroll in Job Corps or to a control group without access to Job Corps.

Analyzing the NJCS data, Schochet and colleagues (2008) reported that access to and participation in the Job Corps program increased educational attainment, reduced criminal activity, and increased earnings for several years after the program. Subsequently, a reanalysis of the NJCS data found that a majority of sites "produced positive effects with widely varying magnitudes, and a minority of sites produced modest negative effects" on earnings four years after the randomization (Weiss et al., 2017, p. 13). They reported the estimated average ITT effect on the total annual earnings to be $1,415 (standard error = $358) and the between-site standard deviation of the site-specific ITT effect to be $1,687.

Researchers have noted differential impacts of Job Corps across youth subpopulations (Blanco et al., 2013; Flores-Lagunes et al., 2010; Frumento et al., 2012; Schochet, 2021). Job Corps was found to provide greater benefits to applicants in their early adulthood rather than in late adolescence; and potential gains especially for minority applicants would not materialize if the local unemployment rate was high. A series of multisite causal mediation analyses has further revealed the mediating role of general education and vocational training in transmitting Job Corps program impacts on earnings (Qin et al., 2021; Qin et al., 2019; Qin & Hong, 2017). While the indirect effects via education and training were found highly consistent across sites, notable variation in the direct effect likely reflects uneven quantity and quality of supplementary services across Job Corps centers. However, these analyses did not rule out the potential contribution of confounding ecological factors to the program impact variation across sites. Meanwhile, problematic incidents at individual Job Corps centers have continued to be a subject of contentious public debate about government investment in the program (Thrush, 2018).

Taken together, an important question remains about how to determine the relative effectiveness of a Job Corps center and pinpoint effective practice. Required by Congressional legislations, the U.S. Department of Labor has had a long history of evaluating job training programs.[2] It has instituted incentives including contract renewals and financial rewards linked to a Job Corps center's performance. According to NJCS researchers, about 10 to 15 contracts were terminated and re-competed each year (Burdghardt & Schochet, 2001; Johnson et al., 1999; Schochet & Burghardt, 2008). However, if the "effectiveness" of a center largely reflects its client composition or is partly determined by the presence of alternative programs or job market conditions in a locality, a Job Corps center would be rewarded or penalized for conditions that are beyond its control; And little is learned about what aspect to improve in its daily practice.

---

[2] The Job Training Partnership Act of 1982 was the first to institute performance measures and accompanying standards for a national workforce program (Blalock & Barnow, 2001). The Workforce Investment Act of 1998 also required performance measures and accountability systems (Fortson et al., 2017).

**Mathematical Definition of Organizational Effectiveness**

To address the preceding issues, this project proposes a rigorous approach to evaluating organizational effectiveness by developing indices less susceptible to the influences of varying local conditions. Consider a population of geographically dispersed sites, each with an organization that operates independently and no meaningful interactions between them. At each site, a Job Corps center serves a local population of eligible clients defined a priori irrespective of organizational practice. Baseline characteristics of the local eligible population are pretreatment, while characteristics of program participants are considered posttreatment because they may be shaped by recruitment practices. To judge the relative effectiveness of individual Job Corps centers in implementing the program, this study introduces an important conceptual distinction between "global relative effectiveness" and "local relative effectiveness."

To determine an organization's *global relative effectiveness* (GRE), organizations are compared as if they all counterfactually operated under the generic ecological conditions of a "synthetic typical site," a hypothetical composite reflecting average client composition, access to alternative programs, and community context across all sites.

However, the generic ecological conditions of "a synthetic typical site" may bear little resemblance to a focal site (e.g., urban vs. rural; mostly Black or Hispanic vs. all White), rendering it an implausible counterfactual. Since the effectiveness of organizational practices often depends on the local context, GRE would not inform how organizations should tailor interventions to local conditions to better serve their unique clientele (Aldrich & Pfeffer, 1976).

In contrast, the *local relative effectiveness* (LRE) of an organization is determined when its performance is compared to that of other organizations operating under similar local ecological conditions. For example, a Job Corps center in a rural region plagued by an opioid epidemic must address its clients' mental health needs. The effectiveness of this center should be assessed alongside other centers in rural areas facing comparable community challenges.

The conceptual distinction between GRE and LRE is fundamentally important. An organization with high LRE under unique local conditions might not appear as effective under hypothetical generic conditions. Therefore, this study focuses on LRE. Hereafter, the term "organizational effectiveness" refers to LRE only.

**Potential Outcomes**

To clarify the concepts, we develop a formal mathematical definition of LRE that employs the potential outcomes framework (Holland, 1986; Rubin, 1978). Let $Z_{ij}$ denote the treatment assignment for individual $i$ at site $j$ that takes value $z = 1$ if the individual is assigned to a Job Corps center and $z = 0$ if assigned to the control condition. It seems reasonable to assume no interference between individuals from different sites; The stable unit treatment value assumption (SUTVA) (Rubin, 1986) further assumes no interference between individuals at the same site. Therefore, an individual's potential outcome is represented as a function of the individual's treatment assignment and does not depend on the treatment assignments of other individuls. Let $Y_{ij}(1)$ denote individual $i$'s potential outcome (such as earnings four years after randomization) if assigned to the Job Corps center at site $j$ while $Y_{ij}(0)$ denote the individual's potential outcome if assigned to the control condition at the same site. $Y_{ij}(1) - Y_{ij}(0)$ defines the *program* *impact* for the individual at the given site, although only one of these two potential outcomes can be observed.

SUTVA also implies that an individual's potential outcome under a given treatment condition does not depend on the mechanisms of treatment assignment or delivery. However, the version of a treatment condition that one receives may depend on the treatment setting (Hong, 2004, 2015; Hong & Raudenbush, 2013). Therefore, we reason that individual $i$'s potential outcome if assigned to the control condition at site $j$ will likely differ from the potential outcome if assigned to the control condition at site $j'$ if ecological conditions vary between these sites. Even if sites $j$ and $j'$ share equivalent ecological conditions, an individual's potential outcome if attending a Job Corps center at site $j$ may still differ from the potential outcome if attending a Job Corps center at site $j'$ if the two centers vary in effectiveness.

**Additional Notation**

Let $\mu_{0j}$ denote the average potential earnings of the eligible population of individuals at site $j$ in the absence of Job Corps and $\mu_{1j}$ for the average potential earnings if they would counterfactually have all been assigned to Job Corps. The population average potential outcomes associated with the experimental and control conditions at site $j$ are defined as $\mu_{0j} = E\left[Y_{ij}(0)\right]$ and $\mu_{1j} = E\left[Y_{ij}(1)\right]$, respectively, where the expectation is taken over the entire population of individuals at site $j$. Hence the site-specific ITT effect is $\mu_{1j} - \mu_{0j}$.

Let $\mathbf{X}_{ij}$ denote a vector of observed pretreatment covariates and $\mathbf{U}_{ij}$ for a vector of unobserved pretreatment covariates that jointly predict the potential outcomes $Y_{ij}(0)$ and $Y_{ij}(1)$. They are "pretreatment" in the sense that they could not have been affected by an individual's assignment to Job Corps. To simplify the notation, henceforth we omit the subscript "$ij$" in $\mathbf{X}_{ij}$ and $\mathbf{U}_{ij}$ for individual $i$ at site $j$. Without losing generality, let $\mathbf{U}$ be orthogonal to $\mathbf{X}$. Let $\boldsymbol{\Phi}_{\mathbf{X}j}$ and $\mathbf{\Phi}_{\mathbf{U}j}$ denote the parameters such as site means and site variances that define the distribution of **X** and that of **U**, respectively, at site $j$. The joint distribution of **X** and **U** at site $j$ is characterized by a vector of parameters: $\mathbf{\Phi}_j = (\mathbf{\Phi}_{\mathbf{X}j}, \mathbf{\Phi}_{\mathbf{U}j})$. In theory, $\mathbf{\Phi}_j$ completely captures the ecological conditions at site $j$ that are unaffected by Job Corps but are predictive of the site-specific ITT effect. If sites $j$ and $j'$ share similar ecological conditions, then $\mathbf{\Phi}_j = \mathbf{\Phi}_{j'}$.

**Definition of Local Relative Effectiveness: The Causal Estimand**

The LRE of the Job Corps center at site $j$ is defined as the difference between the average potential earnings associated with the assignment to Job Corps at this focal site ($\mu_{1j}$) and the average potential earnings associated with Job Corps across all sites with similar ecological conditions, the latter written as $E_{j'}[\mu_{1j'}|\mathbf{\Phi}_{j'} = \mathbf{\Phi}_j]$ where the expectation is taken over all such sites. Therefore, the theoretical definition of the causal estimand is

$$\theta_j = \mu_{1j} - E_{j'}[\mu_{1j'}|\mathbf{\Phi}_{j'} = \mathbf{\Phi}_j]. \tag{1}$$

$\theta_j$ is negative if the LRE of the Job Corps center at the focal site is below the average and is positive if it is above the average in comparison with Job Corps centers at sites with similar ecological conditions. Our definition of LRE is consistent with the Type B effect, which is purely attributable to malleable practices of an organization under its local constraints. Table 1 provides a glossary that includes the key notations for the definition and identification of LRE.

## Identification of Organizational Effectiveness

In this section, we first illustrate how local ecological conditions can confound the evaluation of an organization's LRE when the analysis relies solely on site-specific ITT effects. We then introduce a set of general theoretical models for the causal estimand, based on which we clarify the assumptions required to identify the LRE.

**The Confounding Role of Local Ecological Conditions**

In a multisite randomized trial, the ITT effect at each site can be identified by contrasting the mean observed outcome between the experimental and control groups. However, the LRE of a Job Corps center cannot be identified through a between-site comparison of the site-specific ITT effects if sites differ in their local ecological conditions.

Figures 1a, 1b, and 1c illustrate this point by contrasting hypothetical sites $j$ and $j'$. Suppose the Job Corps center at site $j$ has advantages in local ecological conditions compared to its counterpart at site $j'$. For example, site $j$ serves relatively more mature youths and enjoys a lower local unemployment rate among other site-level features, some of which might be hard to measure. This results in higher average earnings for the control group at site $j$ than at site $j'$ (see a comparison between the two medium grey columns) as well as higher *expected* earnings for the experimental group at site $j$ than at site $j'$ (see a comparison between the two light grey columns) if both two Job Corps centers were counterfactually operating at their respective reference levels of effectiveness—that is, if $\theta_j = \theta_{j'} = 0$.

Figure 1a shows a hypothetical scenario in which the Job Corps center at site $j$ is less effective than the center at site $j'$ (i.e., $\theta_j < \theta_{j'}$, as shown in a comparison between the two patterned columns). This is disguised by the equal ITT effect across these two sites (i.e., $\mu_{1j} - \mu_{0j} = \mu_{1j'} - \mu_{0j'}$) due to the confounding impact of the pre-existing differences in their local ecological conditions given that $\mu_{0j} \neq \mu_{0j'}$ (when comparing the two medium grey columns) and $\mu_{1j} - \theta_j \neq \mu_{1j'} - \theta_{j'}$ (when comparing the two light grey columns).

Even if the two Job Corps centers were equally effective, the ITT effect might nonetheless depend on local ecological conditions that vary between the sites. As illustrated in Figure 1b, the center at site $j$ and that at site $j'$ do not differ in LRE (i.e., $\theta_j = \theta_{j'}$). This is disguised by the unequal ITT effect, again due to the confounding impact of local ecological conditions given that $\mu_{0j} \neq \mu_{0j'}$ and $\mu_{1j} - \theta_j \neq \mu_{1j'} - \theta_{j'}$.

Figure 1c provides a special example in which $\mu_{0j} = \mu_{0j'}$, while local ecological conditions still differ between the two sites because $\mu_{1j} - \theta_j \neq \mu_{1j'} - \theta_{j'}$. Although the centers at these two sites are equally effective (i.e., $\theta_j = \theta_{j'}$), their ITT effects remain unequal.

**General Theoretical Models**

To formalize the above discussion, again the local ecological conditions at site $j$ is characterized by the joint distribution of observed covariates **X** and unobserved covariates **U**, which is denoted by $\boldsymbol{\Phi}_j$ and predicts the average potential outcome associated with the control condition and that associated with Job Corps at the site. Let $g_0$ and $g_1$ each be a flexible function of $\boldsymbol{\Phi}_j$.

$$\mu_{0j} = g_0(\boldsymbol{\Phi}_j);$$

$$\mu_{1j} = g_1(\boldsymbol{\Phi}_j) + \theta_j.$$

Here $g_1(\boldsymbol{\Phi}_j) = \mu_{1j} - \theta_j$ is the counterfactual value of $\mu_{1j}$ when $\theta_j = 0$ -- that is, if the Job Corps center at site $j$ were, possibly contrary to fact, operating at the average level of effectiveness compared to others under similar local ecological conditions (as shown by the light grey column in Figures 1a, 1b, and 1c). Therefore, $g_1(\boldsymbol{\Phi}_j)$ is equal to $E_{j'}[\mu_{1j'} | \boldsymbol{\Phi}_{j'} = \boldsymbol{\Phi}_j]$, which is the reference value for evaluating the LRE of the Job Corps center at focal site $j$. We then partition the site-specific ITT effect $\mu_{1j} - \mu_{0j}$ into the LRE $\theta_j$ and the predicted site-specific ITT effect $g_1(\boldsymbol{\Phi}_j) - g_0(\boldsymbol{\Phi}_j)$, the latter representing the confounding impact of $\boldsymbol{\Phi}_j$ on the ITT effect:

$$\mu_{1j} - \mu_{0j} = \theta_j + [g_1(\boldsymbol{\Phi}_j) - g_0(\boldsymbol{\Phi}_j)]. \quad (2)$$

If sites $j$ and $j'$ share the same joint distribution of **X** and **U**, then $g_1(\boldsymbol{\Phi}_j) - g_0(\boldsymbol{\Phi}_j) = g_1(\boldsymbol{\Phi}_{j'}) - g_0(\boldsymbol{\Phi}_{j'})$; the between-site difference in the ITT effect $(\mu_{1j} - \mu_{0j}) - (\mu_{1j'} - \mu_{0j'})$ is equal to the between-site difference in the LRE $\theta_j - \theta_{j'}$. Figure 2 shows the case that $\mu_{0j} = \mu_{0j'}$ and $\mu_{1j} - \theta_j = \mu_{1j'} - \theta_{j'}$, which satisfies this condition. Only then will the between-site comparison of the ITT effect provide valid information about the LRE.

More often, the joint distribution of **X** and **U** may differ between the sites. The between-site variation in the predicted ITT effect $g_1(\mathbf{\Phi}_j) - g_0(\mathbf{\Phi}_j)$ reflects the heterogeneity in the program impact that is purely due to between-site differences in local ecological conditions. As shown in Figures 1a, 1b, and 1c, when this is the case, simply comparing the site-specific ITT effects is no longer valid for determining the LRE. Rather, it is necessary to adjust for the between-site differences in local ecological conditions.

**Plausibility of Key Identification Assumptions**

Past research on the transportability or generalizability of treatment effects has typically invoked the strong assumption of between-site exchangeability of program impact conditional on observed individual characteristics **X** (Allcott, 2015; Dahabreh et al, 2019, 2020; Hotz et al., 2005; Stuart et al., 2011; Tipton, 2014; Tipton et al., 2017). This assumes that the conditional average treatment effect $E[Y(1) - Y(0)|\mathbf{X} = \mathbf{x}]$ does not differ between sites for individuals with the same observed covariate values **x**. However, this assumption is hard to satisfy due to the omission of **U** as sites may differ in unmeasured client composition, alternative programs available to control group members, and community context. Adjusting for **X** alone is inadequate, but adjusting for both **X** and **U** is generally infeasible.

To address this major limitation, we base our identification strategy on the theoretical insight that a multisite randomized design supplies key information about local ecological conditions at each site. We argue that the distribution of potential outcomes under the control condition $Y(0)$ serves as a proxy for the joint influence of measured and unmeasured contextual factors in the absence of the intervention. These factors may also predict site-specific ITT effects and need to be adjusted for to identify LRE. This is because individuals at the same site with identical covariate values $\mathbf{x}$ may differ in $Y(0)$ due to differences in unmeasured individual characteristics, while individuals at different sites with the same covariate values $\mathbf{x}$ may differ in $Y(0)$ due to differences in not only unmeasured individual characteristics but also relevant contextual features of their respective sites.

This rationale is reflected in the theoretical model for $Y(0)$ at site $j$: $\mu_{0j} = g_0(\mathbf{\Phi}_j)$, indicating that $Y(0)$ contains information about not just $\mathbf{x}$ but also $\mathbf{u}$. Under SUTVA, there is no spillover from experimental to control group within a site; nor is there spillover between sites. $Y(0)$ remains unaffected by the intervention program and is "pretreatment" in nature despite the timing of its measurement. Since $\mathbf{U}$ is not directly observed, we replace $\mathbf{\Phi}_j = (\mathbf{\Phi}_{\mathbf{X}j}, \mathbf{\Phi}_{\mathbf{U}j})$ with $\mathbf{\Phi}_j^* = (\mathbf{\Phi}_{\mathbf{X}j}, \mathbf{\Phi}_{0j})$, where $\mathbf{\Phi}_{0j}$ denotes a vector of parameters characterizing the distribution of $Y(0)$ at site $j$. For sites $j$ and $j'$, $\mathbf{\Phi}_j^* = \mathbf{\Phi}_{j'}^*$ implies that $\mu_{0j} = \mu_{0j'}$. To assess the LRE of each Job Corps center, we invoke the following two assumptions:

***Assumption 1: Weak Conditions for Site Comparability***

$$\text{If } \mathbf{\Phi}_j^* = \mathbf{\Phi}_{j'}^*, \qquad \text{then } \mu_{1j} - \theta_j = \mu_{1j'} - \theta_{j'}, \qquad \forall\ j' \neq j.$$

Under our theoretical model, $\mu_{1j} - \theta_j = g_1(\mathbf{\Phi}_j)$ and $\mu_{1j'} - \theta_{j'} = g_1(\mathbf{\Phi}_{j'})$. If sites $j$ and $j'$ share the same joint distribution of $\mathbf{X}$ and $\mathbf{U}$, then $\mathbf{\Phi}_j = \mathbf{\Phi}_{j'}$, and $\mu_{1j} - \theta_j = \mu_{1j'} - \theta_{j'}$, which is the case illustrated by Figure 2. Assumption 1 instead states that, if two sites share the same joint distribution of $\mathbf{X}$ and $Y(0)$, then they do not differ in the influences of $\mathbf{X}$ and $\mathbf{U}$ on the mean potential outcome under the experimental condition. In other words, it assumes that adjusting for the distribution of $\mathbf{X}$ and $Y(0)$ is as adequate as adjusting for the distribution of $\mathbf{X}$ and $\mathbf{U}$. While still a strong assumption, it is weaker and more plausible than the conventional exchangeability assumption, which relies solely on conditioning on the distribution of **X**.

However, Assumption 1 may be violated if some elements of **U** differ in distribution between sites and predict $\mu_{1j} - \theta_j$ but not $\mu_{0j}$. For example, sites $j$ and $j'$ may differ in demand for highly skilled labor who would be compensated with higher wages when all else is equal across these sites. If Job Corps and alternative job training opportunities are absent, youth in the target population are mostly unequipped with suitable skills and may not differ between the sites in average earnings under the control condition. Their average earnings under the experimental condition, however, would be higher at site $j'$ where Job Corps graduates benefit from more job opportunities requiring high-level skills and offering better compensation. As a result, even if the Job Corps centers at these two sites are equally effective, the ITT effect at site $j'$ would be higher than that at site $j$.

Figure 1c illustrates this scenario, where $\mu_{0j} = \mu_{0j'}$ but $\mu_{1j} - \theta_j < \mu_{1j'} - \theta_{j'}$, revealing the lack of comparability between the two sites in their local ecological conditions. Here, while ecological conditions appear similar under the control condition, they differ under the experimental condition, resulting in different ITT effects across the two sites despite equal effectiveness of the Job Corps centers. In such cases, adjusting for the joint distribution of **X** and $Y(0)$ fails to remove all confounding. Mathematically, the ITT effect at site $j$ is $\mu_{1j} - \mu_{0j} = \theta_j + \left(\mu_{1j} - \theta_j\right) - \mu_{0j}$ and that at site $j'$ is $\mu_{1j'} - \mu_{0j'} = \theta_{j'} + \left(\mu_{1j'} - \theta_{j'}\right) - \mu_{0j'}$. When $\mu_{0j} = \mu_{0j'}$ but $\mu_{1j} - \theta_j \neq \mu_{1j'} - \theta_{j'}$, the between-site difference in ITT effects is unequal to the between-site difference in LRE: $\left(\mu_{1j} - \mu_{0j}\right) - \left(\mu_{1j'} - \mu_{0j'}\right) \neq \theta_j - \theta_{j'}$.

***Assumption 2. Within-Site Ignorable Treatment Assignment***

$$\mu_{zj} = E\left[Y_{ij} | Z_{ij} = z\right], \quad \text{for } z = 0, 1 \text{ and } \forall j.$$

Assumption 2 states that, within every site, the average observed outcome of the experimental group identifies $\mu_{1j}$ while the average observed outcome of the control group identifies $\mu_{0j}$. It also implies that the observed labor market experiences of the control group, which are attributable to not just the measured but also unmeasured composition of clients and other local conditions, are valid counterfactuals for the experimental group in the absence of Job Corps. This assumption is guaranteed by the multisite randomized design.

***Theorem***

When Assumptions 1 and 2 both hold, the LRE of the organization at site $j$ is identified by

$$\theta_j = \left(E\left[Y_{ij}|Z_{ij} = 1\right] - E\left[Y_{ij}|Z_{ij} = 0\right]\right) - E_{j'}\left(E\left[Y_{ij'}|Z_{ij'} = 1, \mathbf{\Phi}_{j'}^{*} = \mathbf{\Phi}_{j}^{*}\right] - E\left[Y_{ij'}|Z_{ij'} = 0, \mathbf{\Phi}_{j'}^{*} = \mathbf{\Phi}_{j}^{*}\right]\right).$$

Here the first term $E\left[Y_{ij}|Z_{ij} = 1\right] - E\left[Y_{ij}|Z_{ij} = 0\right]$ is the average observed outcome difference between the experimental and control groups at the focal site $j$, which identifies the site-specific ITT effect $\mu_{1j} - \mu_{0j}$ under Assumption 2. Under the same assumption, the inner expectation of the second term $E\left[Y_{ij'}|Z_{ij'} = 1, \mathbf{\Phi}_{j'}^{*} = \mathbf{\Phi}_{j}^{*}\right] - E\left[Y_{ij'}|Z_{ij'} = 0, \mathbf{\Phi}_{j'}^{*} = \mathbf{\Phi}_{j}^{*}\right]$ identifies the ITT effect at site $j'$ that shares the same joint distribution of $\mathbf{X}$ and $Y(0)$ as site $j$; the outer expectation averages over all such comparable sites. Under Assumption 1, this second term further identifies the average ITT effect across all sites sharing the same joint distribution of $\mathbf{X}$ and $\mathbf{U}$ as the focal site, which is defined as $E_{j'}\left[\mu_{1j'} - \mu_{0j'}|\mathbf{\Phi}_{j'} = \mathbf{\Phi}_{j}\right]$. The theorem states that the LRE for a focal site is identified by the difference between these two terms. A formal proof is provided in Appendix A.

## Implementation through Mixed-Effects Models

As defined earlier, $\mathbf{\Phi}_{\mathbf{X}j}$ and $\mathbf{\Phi}_{0j}$ constitute the elements of $\mathbf{\Phi}_j^*$ that define the joint distribution of $\mathbf{X}$ and $Y(0)$ at site $j$. To implement, we first obtain measures of $\mathbf{\Phi}_{\mathbf{X}j}$ by collecting a comprehensive set of observed site-level characteristics including aggregated individual characteristics at the baseline and features of the community context at each site. The analyst may want to focus on site-level covariates that show substantial between-site variation. Measures of $\mathbf{\Phi}_{0j}$ are obtained from the aggregated experiences of the control group individuals at each site.

To estimate the LRE, there are alternative parametric and nonparametric approaches to adjusting for the confounding impacts of $\mathbf{\Phi}_{\mathbf{X}j}$ and $\mathbf{\Phi}_{0j}$ with the goal of comparing the site-specific ITT effects among sites that share similar values of $\mathbf{\Phi}_{\mathbf{X}j}$ and $\mathbf{\Phi}_{0j}$. We propose a two-step mixed-effects (2SME) modeling strategy to predict the ITT effect as a function of $\mathbf{\Phi}_{\mathbf{X}j}$ and $\mathbf{\Phi}_{0j}$. When its model-based assumptions as well as the identification assumptions hold, maximum likelihood estimation is expected to generate a consistent and efficient estimate of the LRE for the organization at each site that is independent of the predicted site-specific ITT effect. We further examine these properties through simulations.

**Adjusted Site-Specific ITT effect as a Random Effect**

Suppose that, in a multisite randomized trial, we have observed data for an infinitely large target population at each site. A theoretical model of the ITT effect for individual $i$ in site $j$ is

$$Y_{ij} = \beta_{0j} + \beta_{1j} Z_{ij} + e_{ij}.$$

Here $\beta_{0j}$ is the mean outcome of the control group at site $j$; $\beta_{1j}$ is the ITT effect at the site. We may model $\beta_{0j}$ and $\beta_{1j}$ each as a flexible function of $\mathbf{\Phi}_{\mathbf{X}j}$. The elements of $\mathbf{\Phi}_{\mathbf{X}j}$ that predict $\beta_{0j}$ may or may not be the same as those that predict $\beta_{1j}$. Suppose that the outcome is normally distributed within each site, an assumption that can be easily relaxed. To ease the presentation, below is a simplified representation of the site-level theoretical models in a linear additive form:

$$\beta_{0j} = \alpha_{00} + \boldsymbol{\alpha}_{01}\boldsymbol{\Phi}_{\mathbf{X}j} + \eta_{0j};$$

$$\beta_{1j} = \alpha_{10} + \boldsymbol{\alpha}_{11}\boldsymbol{\Phi}_{\mathbf{X}j} + \eta_{1j};$$

$$\begin{pmatrix} \eta_{0j} \\ \eta_{1j} \end{pmatrix} \sim N\left(\begin{pmatrix} 0 \\ 0 \end{pmatrix}, \begin{pmatrix} \omega_{00} & \omega_{01} \\ \omega_{10} & \omega_{11} \end{pmatrix}\right). \quad (3)$$

The random intercept $\eta_{0j}$ is the site-specific increment to the predicted control group mean $\alpha_{00} + \boldsymbol{\alpha}_{01}\boldsymbol{\Phi}_{\mathbf{X}j}$ at site $j$; The random slope $\eta_{1j}$ is the site-specific increment to the predicted ITT effect $\alpha_{10} + \boldsymbol{\alpha}_{11}\boldsymbol{\Phi}_{\mathbf{X}j}$ at site $j$. The distribution of these two random effects is assumed to be bivariate normal, an assumption that can be empirically examined with sample data. Importantly, the variance of $\eta_{0j}$ denoted by $\omega_{00}$ represents between-site variation in $\beta_{0j}$ unexplained by $\boldsymbol{\Phi}_{\mathbf{X}j}$. For this reason, $\omega_{00}$ is attributed to unmeasured contextual characteristics captured by the control group mean. The variance of $\eta_{1j}$ denoted by $\omega_{11}$ represents between-site variation in $\beta_{1j}$ unexplained by $\boldsymbol{\Phi}_{\mathbf{X}j}$. When the covariance $\omega_{01}$ is nonzero, $\omega_{11}$ is at least partly explained by the remaining between-site differences in the control group mean outcome represented by $\eta_{0j}$.

We subsequently modify the theoretical model for $\beta_{1j}$ as follows:

$$\beta_{1j} = \gamma_{10} + \boldsymbol{\gamma}_{11}\boldsymbol{\Phi}_{\mathbf{X}j} + \gamma_{12}\eta_{0j} + \upsilon_{1j}, \quad \upsilon_{1j} \sim N(0, \tau_{11}). \quad (4)$$

As shown in Appendix B, the new random effect $\upsilon_{1j}$ is the adjusted site-specific ITT effect independent of $\boldsymbol{\Phi}_{\mathbf{X}j}$ and $\eta_{0j}$ and is unconfounded by $\boldsymbol{\Phi}_{\mathbf{X}j}$ and the control group mean. When the identification assumptions and the model-based assumptions hold, $\upsilon_{1j}$ identifies the LRE of the organization at focal site $j$. Table 2 provides a glossary of the key notation used in presenting the mixed-effects models.

**The 2SME Procedure**

By convention, the measurement of $\boldsymbol{\Phi}_{\mathbf{X}j}$ obtained from the sample at site $j$ may be viewed as fixed. However, the sample mean of the control group outcome, which we denote as $\bar{Y}_{0j}$, contains sampling error that can be potentially consequential. It is well known that, when the population mean of the control group outcome $\mu_{0j}$ and the population average ITT effect $\mu_{1j} - \mu_{0j}$ at site $j$ are given, the sampling error in $\bar{Y}_{0j}$ is expected to be negatively associated with the sampling error in the ordinary least squares (OLS) estimator of the site-specific ITT effect $\bar{Y}_{1j} - \bar{Y}_{0j}$. This is because $cov(\bar{Y}_{0j}, \bar{Y}_{1j} - \bar{Y}_{0j} | \mu_{0j}, \mu_{1j}) = -var(\bar{Y}_{0j} | \mu_{0j}, \mu_{1j}) < 0$. When statistically adjusting for $\bar{Y}_{0j}$, the sampling error is expected to partially offset the adjustment for confounding, especially when the sample size at a site is relatively small, creating an "errors in variables" problem that biases the estimation of key parameters.

To circumvent this issue, we adopt a principled solution by implementing a two-step procedure for estimating the random effect $\upsilon_{1j}$ in model (3), which accounts for the reliability of the sample estimator for the control group mean and that for the site-specific ITT effect. Step 1 estimates the control group mean outcome for each site with adjustment for the observed site-level characteristics; Step 2 then estimates the ITT effect for each site with adjustment for the site-specific control group mean and the observed site-level characteristics. Each step employs a mixed-effects model with individuals nested within sites.

***Step 1***. Use the control group data to obtain an empirical Bayes (EB) estimate of the control group mean outcome at site $j$ for $j = 1, \cdots, J$. The EB estimator is a shrinkage estimator that is pulled toward zero in proportion to the amount of sampling noise (Raudenbush & Bryk, 2002). For individual $i$ in site $j$,

$$Y_{ij} = \alpha_{00} + \boldsymbol{\alpha}_{01}\boldsymbol{\Phi}_{\mathbf{X}j} + \eta_{0j} + e_{0ij}; \qquad e_{0ij} \sim N(0, \sigma_0^2); \qquad \eta_{0j} \sim N(0, \omega_{00}).$$

Here $\sigma_0^2$ is the within-site variance of the outcome in the control group; $\eta_{0j}$ represents the component of the control group mean outcome at site $j$ that is independent of $\boldsymbol{\Phi}_{\mathbf{X}j}$. Appendix C provides technical details about obtaining the posterior mean of $\eta_{0j}$ for $j = 1, \cdots, J$.

***Step 2***. Adjust for $\eta_{0j}$ along with $\mathbf{\Phi}_{\mathbf{X}j}$ in analyzing a mixed-effects model for obtaining the adjusted EB estimate of the ITT effect at site $j$.

$$Y_{ij} = \gamma_{00} + \gamma_{01}\mathbf{\Phi}_{\mathbf{X}j} + \gamma_{02}\eta_{0j} + \upsilon_{0j} + \left(\gamma_{10} + \boldsymbol{\gamma}_{11}\mathbf{\Phi}_{\mathbf{X}j} + \gamma_{12}\eta_{0j} + \upsilon_{1j}\right)Z_{ij} + e_{ij};$$

$$e_{ij} \sim N\left(0, Z_{ij}\sigma_1^2 + \left(1 - Z_{ij}\right)\sigma_0^2\right); \qquad \begin{pmatrix}\upsilon_{0j}\\ \upsilon_{1j}\end{pmatrix} \sim N\left(\begin{pmatrix}0\\0\end{pmatrix}, \mathbf{T}\right) \text{ where } \mathbf{T} = \begin{pmatrix}\tau_{00} & \tau_{01}\\ \tau_{10} & \tau_{11}\end{pmatrix}.$$

Here $\sigma_1^2$ is the within-site variance of the outcome in the experimental group. Conditioning on $\mathbf{\Phi}_{\mathbf{X}j}$ and $\eta_{0j}$, no remaining between-site variation in the control group mean is expected, so $\upsilon_{0j} = 0$ and $\tau_{00} = \tau_{01} = 0$ in theory. The predicted ITT effect at site $j$ is a function of $\mathbf{\Phi}_{\mathbf{X}j}$ and $\eta_{0j}$ and reflects the confounding influence of local ecological conditions as captured by the joint distribution of $\mathbf{X}$ and $Y(0)$. As shown in Equation (4), $\upsilon_{1j}$ is the discrepancy between the ITT effect and the predicted ITT effect at site $j$; $\upsilon_{1j}$ identifies the LRE ($\theta_j$) that is independent of confounding factors when the identification and model-based assumptions hold. In practice, $\eta_{0j}$ is replaced by $\eta_{0j}^*$ obtained from step 1. Appendix C provides technical details for obtaining the posterior mean of $\upsilon_{1j}$, denoted as $\upsilon_{1j}^*$, as a sample estimator of $\theta_j$, additionally accounting for the reliability of the site-specific ITT effect estimator. Maximum likelihood estimation can be conducted with standard software such as Stata and R for mixed-effects models.

## Simulations

To evaluate the performance of our proposed strategy, we conduct Monte Carlo simulations to address two questions: (1) How does the performance of the 2SME procedure compare with alternative strategies when Assumption 1 is only mildly violated? (2) How do these strategies compare when Assumption 1 is severely violated? We make comparisons primarily between the following four candidate strategies for estimating LRE.

- *2SME*: a two-step mixed-effects modeling procedure adjusting for $\mathbf{\Phi}_{\mathbf{X}j}$ and $\mathbf{\Phi}_{0j}$, which is our proposed strategy;
- *Mean-centered ITT*: a site-by-site OLS analysis of the site-specific increment to the average ITT effect;
- *ME_adj_X*: a one-step mixed-effects model adjusting for $\mathbf{\Phi}_{\mathbf{X}j}$ only;
- *ME_adj_X_U*: a one-step mixed-effects model adjusting for $\mathbf{\Phi}_{\mathbf{X}j}$ and $\mathbf{\Phi}_{\mathbf{U}j}$.

In all cases, $\mathbf{\Phi}_{\mathbf{X}j}$, $\mathbf{\Phi}_{\mathbf{U}j}$, and $\mathbf{\Phi}_{0j}$ are replaced by their sample analogues such as the estimated site means of $\mathbf{X}$, $\mathbf{U}$, and the observed outcome of the control group, respectively. Although the last model *ME_adj_X_U* is infeasible in practice, we include it in our simulations to provide a benchmark since it does not rely on Assumption 1. While the second model *Mean-centered ITT* is a different parameter than the LRE, the former can be considered a naïve estimator of the latter. As such, we include it in the comparison.

To address research question (1), we simulate a scenario where the violation of Assumption 1 is relatively minor. Research question (2) is addressed through a scenario with a severe violation of Assumption 1. Conventional evaluation criteria include bias, efficiency, and root mean square error (RMSE) for the point estimator of LRE at each site. However, since LRE involves as many as $J$ parameters of interest, we also consider severe classification error (SCE) and moderate classification error (MCE) rates. The Job Corps performance measurement system classifies centers into three tiers based on their performance—top 30%, middle 40%, and bottom 30% (Schochet & Burghardt, 2008). A severe classification error occurs when a center in the top tier based on the true value of its LRE is misclassified into the bottom tier, or vice versa; a moderate classification error occurs when a center in the top or bottom tier is misclassified into the middle tier, or vice versa.

According to the simulation results, we conclude that among the three feasible candidate strategies under consideration, the *2SME* model performs best. In other words, the 2SME estimator is least biased, most accurate, and makes the fewest severe classification errors.

(1) Under scenario 1, the *2SME* model removes nearly as much bias as the infeasible model that adjusts for both observed and unobserved site-level covariates. In contrast, adjusting for only observed covariates retains bias from unobserved covariates, and unadjusted ITT analysis shows the greatest bias. The *2SME* estimator also demonstrates higher efficiency and accuracy and lower SCE and MCE rates compared to the alternative strategies.

(2) Under scenario 2, although the *2SME* model generally underperforms the infeasible model, it consistently outperforms other feasible candidate strategies including the *ITT* model and the *ME_adj_X* model in bias removal, efficiency, accuracy, and SCE and MCE rates.

The derivation in Appendix C indicates that our LRE estimator is consistent when the assumptions are satisfied. Simulation results under scenario 1 show that both the average absolute bias across sites and the between-site standard deviation of bias decrease as the number of sites and the sample size per site simultaneously increase. Details on data generation, evaluation criteria, and simulation results are provided in the supplementary material.

## Re-analysis of the NJCS Data

We apply the 2SME strategy to estimate the LRE of Job Corps centers using the NJCS data. The sample consists of 10,001 individuals with valid site membership and outcome data: self-reported weekly earnings from the 48-month post-randomization survey. Of these, 6,072 were assigned at random to Job Corps and 3,929 to the control condition. The sample size per site ranged from 27 to 489, with treatment shares per site between 48% and 75%. Four years after randomization, applicants assigned to Job Corps earned about $13 more per week (roughly $28 in 2025 dollars) than those in the control group on average (standard error = \$4.7), representing a roughly seven percent increase in earnings.[3] The estimated between-site standard deviation of the ITT effect is about \$23.[4] The top section of Table 3 presents sample information and outcome distributions across sites.

**Covariate Selection and Model Specification**

We consider two sets of site-level covariates. The first set comprises a minimal selection of theoretically important individual-level measures, aggregated to the site level, to describe the local ecological conditions at each Job Corps center. These include demographic compositions (age, gender, and race/ethnicity) as well as earnings and employment history. As shown in Table 3, the sites differ considerably in the distribution of each of these covariates. The second set includes a broader range of potentially relevant site-level covariates.[5] To maximize confounding removal while preventing model overfitting, we apply the least absolute shrinkage and selection operator (LASSO) with the recommended "1-standard error" penalty (Hastie et al., 2015; Hastie et al., 2017) to select covariates from this second set. The results indicate that adjusting for only the first set of covariates is adequate.

To implement the 2SME strategy, we analyze the control group data in Step 1, regressing Year 4 earnings on the first set of covariates via a mixed-effects model with a random intercept and saving the EB residual of the control group mean outcome for each site. We then combine the control group and experimental group data in Step 2, regressing earnings on the treatment, the first set of covariates, the EB residual of the control group mean outcome obtained from Step 1, and the interaction between the treatment and each of these covariates. The Step 2 mixed-effects model includes a random slope for the treatment as well as a random intercept. We obtain the EB residual of the random slope as the LRE estimate for the Job Corps center at each site.

---

[3] We adjusted for inflation using the U.S. Bureau of Labor Statistic's CPI inflation calculator: https://www.bls.gov/data/inflation_calculator.htm

[4] We compared the results across three different mixed-effects models: (1) a random-intercept random-slope model with an uncentered treatment indicator; (2) a random-intercept random-slope model with a group-mean centered treatment indicator; and (3) a fixed-intercept random-slope model. All three models produced very similar maximum likelihood estimates of the average ITT effect and the between-site standard deviation of the ITT effect, which indicates that the site-specific ITT effect is uncorrelated with the precision of its estimation at each site in this study.

[5] The second set of covariates includes public benefit receipt, education, health, English language learner status, criminal history, household size at the baseline as well as application location and timing, population density of residential area, and within-site proportion of applicants with nonresidential designation status.

**Summary of Analytic Results**

This section presents the results of applying the 2SME procedure to the NJCS data, emphasizing the role of the control group mean outcome in addressing unmeasured confounding. We visualize the distribution of estimated LREs independent of the predicted site-specific ITT effects, the latter of which capture pre-existing differences in ecological conditions between sites. Finally, we highlight changes in center rankings when Job Corps centers are evaluated using LRE estimates rather than site-specific ITT effect estimates.

***Significance of Adjustment for Control Group Mean Outcome***. Table 4 presents the estimation results from the 2SME specification. After accounting for the first set of site-level covariates, the EB residual of the control group mean outcome emerges as a significant additional predictor of the site-specific ITT effect. This finding underscores the importance of controlling for confounding information reflected in the control group mean outcome, which is not captured by the observed baseline covariates.

***Between-Site Variation in LRE***. Once these sources of confounding are accounted for, the estimated between-site standard deviation of the LRE is $21. After adjusting for local ecological conditions, the expected outcome for an organization based on its relative effectiveness is likely to fall between approximately $40 below to $40 above the conditional average. Figure 3 presents a scatterplot that decomposes between-site variation in the ITT effect

into two distinct dimensions. The horizontal axis reflects variation in the predicted ITT effect as a function of the observed site-level baseline covariates and the EB residual of the control group mean outcome. Sites with similar predicted ITT effects generally share comparable local ecological conditions, as captured by these contextual features. The vertical axis shows the between-site variation in the estimated LRE at each level of predicted ITT effect. Despite the shrinkage due to the EB procedure, substantial variation remains in the estimated LRE among sites with similar ecological conditions. This pattern indicates that even among Job Corps centers operating under similar circumstances, organizational effectiveness varies considerably.[6]

***Comparison of Job Corps Center Rankings: ITT Effect vs. LRE***. Accounting for pre-existing between-site differences in local ecological conditions has led to a reordering of some of the organizations previously identified as among the "best" or "worst." A comparison of the rankings of Job Corps centers based on the 2SME estimates of the LRE with those based on the EB estimates of the unadjusted ITT effect reveals notable discrepancies between the two across many sites. A visual depiction of this evidence can be found in the online supplementary material. The estimated ITT effect and the estimated LRE are only moderately correlated ($r$ = 0.62). Notably, several Job Corps centers that rank highly in terms of the unadjusted ITT effect show a negative LRE estimate representing a below-average level of relative effectiveness.

## Conclusions and Discussion

In a multisite randomized trial, impact variation across sites could be due to uneven effectiveness in implementation across the organizations or due to differences in ecological conditions across the community contexts. Disentangling the former from the latter will enable researchers to investigate theoretically significant questions such as what makes an organization effective under a set of given local conditions (Patton, 2001). Invalid assessments of organizational effectiveness could mislead further investigations of desirable organizational practice.

---

[6] Our simulations indicate that neither the posterior variance nor the bootstrap variance reliably quantifies the estimation uncertainty of the LRE estimates produced by the 2SME procedure. Therefore, we refrain from making definitive assessments regarding the statistical significance of differences in LRE between any two sites.

**Contributions of This Study**

This project presents a methodological advance in the initial phase of this research program. It focuses on bringing clarity to scientific discussions around organizational effectiveness through an explicit mathematical definition of local relative effectiveness. The LRE estimand contrasts the effectiveness of a focal organization with others that share its local ecological conditions. Leveraging multisite randomized trials, we develop an innovative identification strategy that incorporates not just baseline information from all individuals but also post-randomization data from control group members at each site, the latter capturing part of otherwise unmeasured pre-existing between-site differences. As a result, our method successfully disentangles organizational effectiveness from both observed local community features that directly predict heterogeneity in ITT effects and unobserved site-level characteristics that influence mean outcomes under the control condition and thereby indirectly contribute to ITT heterogeneity. In general, adjusting for the control group mean outcome is particularly important in applications where baseline site-level characteristics are inadequately measured, a common scenario in randomized trials especially when data on lagged outcomes are lacking.

As shown in our simulation results, because Assumption 1 is considerably more plausible than the conventional strong assumption of site exchangeability that conditions on observed baseline site-level covariates only, our new evaluation scheme generates LRE indices less biased than the conventional ones. Moreover, there is considerable gain in efficiency as well as in the overall accuracy of LRE estimation, which leads to a substantial reduction in classification error rates even when Assumption 1 is severely violated.

The application to the NJCS data demonstrates the feasibility of obtaining LRE estimates by adjusting for the EB estimates of the control group average outcome at each site along with other site-level baseline covariates. In this empirical study, the former is an additional significant predictor of the site-specific ITT effect even after statistical adjustment has been made for the observed baseline covariates. The results reveal notable discrepancies between the unadjusted ITT effects and the LRE estimates, underscoring the distinction between questions about the ITT effects and those about LRE. For example, some Job Corps centers seem to exhibit a positive ITT effect but a zero or negative LRE. While these centers may offer valuable services that would otherwise be unavailable to eligible clients without Job Corps, their value-added could potentially be enhanced by adopting the practices of more effective peer organizations.

There has been an increase of multisite randomized trials or multisite natural experiments for evaluating programs in fields as diverse as education, training, healthcare, and social services. The feasibility of treatment randomization (or a natural experiment such as a lottery) may depend on whether an organization is oversubscribed at a site. Nonetheless, when such research opportunities present themselves, methodological advances enable valid evaluations of organizational effectiveness, which paves the way for further theoretical investigations with broad implications for organizational improvement.

**Implications for Future Research on Organizational Effectiveness**

The insights and methods developed in this study provide a strong foundation for empirical research on predictors of organizational effectiveness, program contextualization, analysis of diverse outcomes, and other applications for testing program theories (Reynolds, 1998). Further extensions of these methodological tools will enable researchers to generate valuable contextualized knowledge about organizational effectiveness that complements and enriches the existing literature.

***Investigating Predictors of Organizational Effectiveness.*** The empirical results suggest that Job Corps centers were indeed largely uneven in their effectiveness. In theory, organizations may enhance their practices by replacing leadership, training employees, and mobilizing or reallocating resources. Structural features of Job Corps centers such as center size (Burghardt & Schochet, 2001), student-to-counselor ratios (Johnson et al, 1995), and the quantity and quality of supplementary service provisions (Qin, Deutsch, & Hong, 2021) are additional malleable factors. With site-level LRE as the outcome of interest, evaluation researchers may investigate how these organizational characteristics, either independently or in combination, predict the relative effectiveness of Job Corps centers.

***Tailoring Interventions to Local Contexts.*** The proposed strategy may be extended to investigate how intervention programs can be optimally tailored to local contexts. Certain organizational practices may influence LRE differently depending on local ecological conditions. For example, urban and rural centers may require different types of practices to achieve effectiveness. In communities facing severe resource deprivation, investments in supplementary services may be essential for centers to effectively serve youth.

***Examining Varied Outcomes.*** An organization may not demonstrate equal effectiveness across different outcome domains. In the current study, we evaluate organizational effectiveness in terms of clients' medium-term earnings. Given the Job Corps program's whole-person approach, future research may examine additional outcomes such as job placement, educational attainment, improvements in social skills and mental health, reductions in victimization and criminal involvement, and long-term earnings (Schochet, 2021). One may further explore combining these outcomes into low-dimensional factors or creating a composite index to provide a more holistic assessment of organizational effectiveness.

***Applying to Multisite Evaluation.*** Future research may explore extensions of our strategy beyond multisite randomized trials to multisite natural experiments including matched pairs or stratified cluster randomized trials (Imai, King, & Nall, 2009; Turner et al, 2017). In these alternative designs, organizations are matched or stratified on baseline characteristics. Each matched pair or stratum forms a pseudo site, within which organizations are as if randomized to either the experimental or control condition. Potentially for all these cases, our methodological advancements have increased the feasibility of developing an index of organizational effectiveness with greater conceptual clarity and enhanced validity compared to conventional approaches. However, we caution that when the number of sites or the sample size per site are overly small, there are legitimate concerns about the accuracy of estimation. Nonetheless, even imprecise and inaccurate teacher value added measures have been found useful for teacher evaluation (Staiger & Kane, 2015).

**Limitations and Further Methodological Inquiry**

Given the novelty of the proposed 2SME strategy, especially given that the causal parameters of interest are as many as the number of organizations in a study, we highlight several methodological questions for further investigation.

***Statistical Inference***. Past research has shown that, for high-dimensional inference, the under-coverage rate of parametric empirical Bayes confidence intervals increases with the degree of shrinkage (Armstrong et al., 2022). Indeed, our simulations reveal that the estimated posterior variance of the LRE estimate tends to diverge from the empirical variance, especially when the between-site standard deviation of the LRE is limited and when the sample size is relatively small, both contributing factors to the magnitude of shrinkage. Our simulations also reveal notable divergence of bootstrap variance from the empirical variance. Furthermore, the use of standard confidence intervals has been found invalid for analyses employing machine learning methods such as LASSO (Efron & Hastie, 2016). We will investigate in future research the performance of credible intervals for the LRE obtained via Bayesian inference.

***Model Specification***. The results from the 2SME approach may be sensitive to model misspecification. To address this, one may consider alternative semiparametric or nonparametric approaches. Future research may extend Bayesian Additive Regression Trees (BART; Chipman et al., 2010) or Bayesian Causal Forest (BCF; Hahn et al., 2020) methods to multisite data. With high-dimensional covariates, past research has proposed cross-fitting techniques to reduce overfitting bias when a single structural or causal parameter is of policy interest (Chernozhukov et al., 2018). Future research may adapt the cross-fitting approach to settings involving high-dimensional causal parameters and investigate its impact on the statistical properties of the LRE estimator.

***Sensitivity Analysis***. Our proposed strategy relies on Assumption 1, which may be violated if omitted site-level covariates predict the site-specific ITT effect but not the site-specific mean outcome of the control group. To assess the potential consequence of omitted confounding, future research may extend simulation-based sensitivity analysis methods (Dorie et al., 2016; Hong et al., 2023; Qin & Yang, 2022) by simulating plausible distributions of omitted covariates. The initial evaluation results are sensitive to the omission if the sensitivity analysis yields qualitatively different conclusions about the LRE for some organizations.

***Noncompliance***. The current study has focused on the potential outcome of the treatment assigned rather than the treatment received. This is suitable because, with control group members having no access to Job Corps, a higher level of compliance rate at a site indicates a higher level of effectiveness of the Job Corps center in recruiting and retaining eligible applicants assigned to the experimental group. If the focus is instead on evaluating organizational effectiveness in serving clients who participate in a program, the potential outcome of the treatment received may become more relevant. In randomized experiments with noncompliance, the instrumental variable (IV) method has been proposed (Bloom, 1984) to identify and estimate the complier average treatment effect (CATE) (Angrist et al., 1996). Future extensions may incorporate the IV method by replacing the site-specific ITT effect with the site-specific CATE and additionally explore ways to address partial compliance among dropouts (Heckman et al., 1998).

***Potential Spillover***. By convention, we assume that the experimental group and the control group are independent random samples of the eligible population at each site and that the presence of the experimental condition would not change the general equilibrium in the local labor market. These assumptions may not always hold. Moreover, in some multisite trials, the same organization may serve both the experimental group and the control group. Such a design would increase the likelihood of spillover from one treatment condition to the other, complicating the identification of LRE. We caution against directly applying our proposed method to these cases.

**Acknowledgements**

The research reported here was supported by the Institute of Education Sciences (IES), U.S. Department of Education, through a Statistical and Research Methodology Grant (R305D120020) and a major research grant from the Spencer Foundation (202400131). The opinions expressed are those of the authors and do not represent views of the funding agencies. The authors would like to thank Howard Bloom, Jake Bowers, Avi Feller, George Karabatsos, Xinran Li, Laura Peck, Peter Schochet, Jeff Smith, Stephen Raudenbush, participants at the University of Chicago Education Workshop and the University of Illinois Chicago Statistics and Data Science Seminar, and four anonymous reviewers for their valuable suggestions.

# Appendix A

## Proof of the Theorem

We prove that when Assumptions 1 and 2 hold, the difference in the average ITT effect between a focal site and the comparable sites that share the same joint distribution of $\mathbf{X}$ and $Y(0)$ identifies the LRE of the organization at the focal site. Under Assumption 2,

$$\left(E\left[Y_{ij}|Z_{ij}=1\right]-E\left[Y_{ij}|Z_{ij}=0\right]\right)$$

$$-E_{j'}\left(E\left[Y_{ij'}|Z_{ij'}=1,\mathbf{\Phi}_{j'}^{*}=\mathbf{\Phi}_{j}^{*}\right]-E\left[Y_{ij'}|Z_{ij'}=0,\mathbf{\Phi}_{j'}^{*}=\mathbf{\Phi}_{j}^{*}\right]\right)$$

$$=\left[\mu_{1j}-\mu_{0j}\right]-E_{j'}\left[\mu_{1j'}-\mu_{0j'}|\mathbf{\Phi}_{j'}^{*}=\mathbf{\Phi}_{j}^{*}\right]$$

$$=\left[\theta_{j}+\mu_{1j}-\theta_{j}-\mu_{0j}\right]-E_{j'}\left[\theta_{j'}+\mu_{1j'}-\theta_{j'}-\mu_{0j'}|\mathbf{\Phi}_{j'}^{*}=\mathbf{\Phi}_{j}^{*}\right]$$

$$=\left(\theta_{j}-E_{j'}\left[\theta_{j'}|\mathbf{\Phi}_{j'}^{*}=\mathbf{\Phi}_{j}^{*}\right]\right)+\left(\left[\mu_{1j}-\theta_{j}\right]-E_{j'}\left[\mu_{1j'}-\theta_{j'}|\mathbf{\Phi}_{j'}^{*}=\mathbf{\Phi}_{j}^{*}\right]\right)-\left(\mu_{0j}-E_{j'}\left[\mu_{0j'}|\mathbf{\Phi}_{j'}^{*}=\mathbf{\Phi}_{j}^{*}\right]\right).$$

When $\mathbf{\Phi}_{j'}^{*} = \mathbf{\Phi}_{j}^{*}$, by definition, we have that $\mu_{0j} = \mu_{0j'} = E_{j'}\left[\mu_{0j'}|\mathbf{\Phi}_{j'}^{*} = \mathbf{\Phi}_{j}^{*}\right]$. Furthermore, under Assumption 1, when $\mathbf{\Phi}_{j'}^{*} = \mathbf{\Phi}_{j}^{*}$, we have that $\mu_{1j} - \theta_j = \mu_{1j'} - \theta_{j'} = E_{j'}\left[\mu_{1j'} - \theta_{j'}|\mathbf{\Phi}_{j'}^{*} = \mathbf{\Phi}_{j}^{*}\right]$. Hence the above is equal to

$$\theta_j - E_{j'}\left[\theta_{j'}|\mathbf{\Phi}_{j'}^{*} = \mathbf{\Phi}_{j}^{*}\right]$$

$$= \theta_j.$$

The last equation holds because, by definition, $E_{j'}\left[\theta_{j'}|\mathbf{\Phi}_{j'}^{*} = \mathbf{\Phi}_{j}^{*}\right] = 0$ is the reference value of LRE for the organization at the focal site $j$.

## Appendix B

## LRE Estimator Unconfounded by Control Group Mean Outcome and Observed Baseline Covariates

Here we show that the random effect $v_{1j}$ in model (4) is equal to the difference between the site-specific ITT effect and the predicted ITT effect at focal site $j$, the latter being a function of the control group mean outcome as well as $\mathbf{\Phi}_{\mathbf{X}j}$. Given the site-level theoretical models specified in (3), the joint distribution of $\beta_{0j}$ and $\beta_{1j}$ is bivariate normal:

$$\begin{pmatrix} \beta_{0j} \\ \beta_{1j} \end{pmatrix} | \mathbf{\Phi}_{\mathbf{X}j} \sim N\left(\begin{pmatrix} \alpha_{00} + \boldsymbol{\alpha}_{01}\mathbf{\Phi}_{\mathbf{X}j} \\ \alpha_{10} + \boldsymbol{\alpha}_{11}\mathbf{\Phi}_{\mathbf{X}j} \end{pmatrix}, \begin{pmatrix} \omega_{00} & \omega_{01} \\ \omega_{10} & \omega_{11} \end{pmatrix}\right).$$

Since $\eta_{0j} = \beta_{0j} - \alpha_{00} - \boldsymbol{\alpha}_{01}\mathbf{\Phi}_{\mathbf{X}j}$, we can derive the conditional mean of $\beta_{1j}$ as follows:

$$E\left[\beta_{1j}|\beta_{0j}, \mathbf{\Phi}_{\mathbf{X}j}\right] = \alpha_{10} + \boldsymbol{\alpha}_{11}\mathbf{\Phi}_{\mathbf{X}j} + \frac{\omega_{01}}{\omega_{00}}\left(\beta_{0j} - \alpha_{00} - \boldsymbol{\alpha}_{01}\mathbf{\Phi}_{\mathbf{X}j}\right)$$

$$= \left(\alpha_{10} - \frac{\omega_{01}}{\omega_{00}}\alpha_{00}\right) + \left(\boldsymbol{\alpha}_{11} - \frac{\omega_{01}}{\omega_{00}}\boldsymbol{\alpha}_{01}\right)\mathbf{\Phi}_{\mathbf{X}j} + \frac{\omega_{01}}{\omega_{00}}\eta_{0j}.$$

Let $\gamma_{10} = \alpha_{10} - \frac{\omega_{01}}{\omega_{00}}\alpha_{00}$, $\boldsymbol{\gamma}_{11} = \boldsymbol{\alpha}_{11} - \frac{\omega_{01}}{\omega_{00}}\boldsymbol{\alpha}_{01}$, $\gamma_{12} = \frac{\omega_{01}}{\omega_{00}}$, and $\tau_{11} = \left(1 - \frac{\omega_{01}^2}{\omega_{00}\omega_{11}}\right)\omega_{11}$.

We then represent the predicted ITT effect at site $j$ in a model that reflects the confounding influences of the control group mean outcome as well as the observed baseline covariates:

$$E\left[\beta_{1j}|\beta_{0j}, \boldsymbol{\Phi}_{\mathbf{X}j}\right] = \gamma_{10} + \boldsymbol{\gamma}_{11}\boldsymbol{\Phi}_{\mathbf{X}j} + \gamma_{12}\eta_{0j}.$$

The random effect $\upsilon_{1j}$ in model (4) is the increment to this predicted site-specific ITT effect and is equal to $\beta_{1j} - \gamma_{10} - \boldsymbol{\gamma}_{11}\boldsymbol{\Phi}_{\mathbf{X}j} - \gamma_{12}\eta_{0j}$. We can derive

$E\left[\upsilon_{1j}|\beta_{0j}, \boldsymbol{\Phi}_{\mathbf{X}j}\right] = E\left[\beta_{1j} - \gamma_{10} - \boldsymbol{\gamma}_{11}\boldsymbol{\Phi}_{\mathbf{X}j} - \gamma_{12}\eta_{0j}|\beta_{0j}, \boldsymbol{\Phi}_{\mathbf{X}j}\right]$

$= E\left[\beta_{1j}|\beta_{0j}, \boldsymbol{\Phi}_{\mathbf{X}j}\right] - \left(\gamma_{10} + \boldsymbol{\gamma}_{11}\boldsymbol{\Phi}_{\mathbf{X}j} + \gamma_{12}\eta_{0j}\right)$

$= 0.$

Therefore, $\upsilon_{1j}$ is independent of $\beta_{0j}$ and $\boldsymbol{\Phi}_{\mathbf{X}j}$ and identifies the LRE of the organization at site $j$ under Assumptions 1 and 2 given that the confounding influences of the control group mean as well as the observed baseline covariates have been removed from the site-specific ITT effect.

## Appendix C

## Empirical Bayes Estimator of LRE

Although we view $\boldsymbol{\Phi}_{\mathbf{X}j}$ as fixed for site $j$, the control group mean outcome $\beta_{0j}$ and the ITT effect $\beta_{1j}$ at the site are to be estimated from the sample data.

In step-1 analysis, the variance of the site-specific random intercept $\eta_{0j}$ captures the between-site variation in the control group mean outcome that is unexplained by $\boldsymbol{\Phi}_{\mathbf{X}j}$ and is attributable to some elements of $\boldsymbol{\Phi}_{\mathbf{U}j}$. The posterior mean of $\eta_{0j}$ is $E\left[\eta_{0j}|Y_{ij}, \boldsymbol{\Phi}_{\mathbf{X}j}, \alpha_{00}, \boldsymbol{\alpha}_{01}, \sigma_0^2, \omega_{00}\right] = \lambda_{0j}\left(\bar{Y}_{0j} - \alpha_{00} - \boldsymbol{\alpha}_{01}\boldsymbol{\Phi}_{\mathbf{X}j}\right)$ and the posterior variance $\omega_{00}\left(1 - \lambda_{0j}\right)$. Here $\lambda_{0j} = \omega_{00}/\left(\omega_{00} + \sigma_0^2/n_{0j}\right)$ is the reliability of $\bar{Y}_{0j} - \alpha_{00} - \boldsymbol{\alpha}_{01}\boldsymbol{\Phi}_{\mathbf{X}j}$ as an estimator of $\eta_{0j}$, with $n_{0j}$ denoting the sample size of the control group in site $j$ (Raudenbush & Bryk, 2002). The reliability is lower when $n_{0j}$ is smaller, when $\sigma_0^2$ is greater, or when $\omega_{00}$ is smaller. The EB estimator of $\eta_{0j}$, denoted as $\eta_{0j}^*$, is a shrinkage estimator and is equal to $\hat{\lambda}_{0j}(\bar{Y}_{0j} - \hat{\alpha}_{00} - \hat{\boldsymbol{\alpha}}_{01}\boldsymbol{\Phi}_{\mathbf{X}j})$. The lower the reliability, the greater the shrinkage of $\eta_{0j}^*$ toward zero.

In step-2 analysis, the reliability matrix for the sample estimator of the control group mean outcome $\bar{Y}_{0j}$ and that of the site-specific ITT effect $\bar{Y}_{1j} - \bar{Y}_{0j}$ is

$$\boldsymbol{\Lambda}_j = \mathbf{T}(\mathbf{T} + \mathbf{V}_j)^{-1} = \begin{pmatrix} \lambda_{0j} & \lambda_{01j} \\ \lambda_{10j} & \lambda_{11j} \end{pmatrix},$$

$$\text{where } \mathbf{V}_j = var\begin{pmatrix} \bar{Y}_{0j} \\ \bar{Y}_{1j} - \bar{Y}_{0j} \end{pmatrix} = \begin{pmatrix} \sigma_0^2/n_{0j} & -\sigma_0^2/n_{0j} \\ -\sigma_0^2/n_{0j} & \sigma_1^2/n_{1j} + \sigma_0^2/n_{0j} \end{pmatrix}.$$

Also let $\boldsymbol{\Gamma} = (\gamma_{00}, \boldsymbol{\gamma}_{01}, \gamma_{02}, \gamma_{10}, \boldsymbol{\gamma}_{11}, \gamma_{12})$ and $\boldsymbol{\Sigma} = (\sigma_1^2, \sigma_0^2, \tau_{00}, \tau_{10}, \tau_{11})$. The posterior mean of $\upsilon_{1j}$ is

$$\begin{aligned} &E[\upsilon_{1j} | Y_{ij}, Z_{ij}, \boldsymbol{\Phi}_{\mathbf{X}j}, \eta_{0j}, \boldsymbol{\Gamma}, \boldsymbol{\Sigma}] \\ &\qquad = \lambda_{10j}(\bar{Y}_{0j} - \gamma_{00} - \boldsymbol{\gamma}_{01}\boldsymbol{\Phi}_{\mathbf{X}j} - \gamma_{02}\eta_{0j}) \\ &\qquad + \lambda_{11j}[(\bar{Y}_{1j} - \bar{Y}_{0j}) - (\gamma_{10} + \boldsymbol{\gamma}_{11}\boldsymbol{\Phi}_{\mathbf{X}j} + \gamma_{12}\eta_{0j})]. \end{aligned}$$

As $\bar{Y}_{0j} - \gamma_{00} - \boldsymbol{\gamma}_{01}\boldsymbol{\Phi}_{\mathbf{X}j} - \gamma_{02}\eta_{0j}$ is equal to zero in expectation, the posterior mean is simplified to be $\lambda_{11j}[(\bar{Y}_{1j} - \bar{Y}_{0j}) - (\gamma_{10} + \boldsymbol{\gamma}_{11}\boldsymbol{\Phi}_{\mathbf{X}j} + \gamma_{12}\eta_{0j})]$. The posterior variance of $\upsilon_{1j}$ is

$$var(\upsilon_{1j} | Y_{ij}, Z_{ij}, \boldsymbol{\Phi}_{\mathbf{X}j}, \eta_{0j}, \boldsymbol{\Gamma}, \boldsymbol{\Sigma}) = -\lambda_{10j}\tau_{10} + (1 - \lambda_{11j})\tau_{11}.$$

Since $\tau_{10}$ is equal to zero in expectation when conditioning on $\boldsymbol{\Phi}_{\mathbf{X}j}$ and $\eta_{0j}$, the posterior variance is simplified to be $(1 - \lambda_{11j})\tau_{11}$. To estimate the posterior variance, the elements of $\boldsymbol{\Gamma}$ and $\boldsymbol{\Sigma}$ are to be replaced by their sample analogues while $\eta_{0j}$ is to be replaced by $\eta_{0j}^*$ obtained from the step-1 analysis.

When the sample size per site goes to infinity, the reliability of the control group mean outcome estimator and that of the site-specific ITT effect estimator will both converge to 1, in which case $\eta_{0j}^{*}$ and $\upsilon_{1j}^{*}$ will converge to the true values of $\eta_{0j}$ and $\upsilon_{1j}$, respectively. Therefore, when the identification assumptions and the model-based assumptions hold, $\upsilon_{1j}^{*}$ is a consistent estimator of $\theta_j$. If the sample size at a site is relatively small or if the between-site variance of LRE denoted by $\tau_{11}$ is relatively small, $\lambda_{11j}$ will be relatively low. In such a case, the EB estimator of the posterior mean denoted by $\upsilon_{1j}^{*}$ will shrink toward 0 while the posterior variance will increase, which will indicate a relatively high level of uncertainty in estimation.

Table 1. Glossary for Defining and Identifying Local Relative Effectiveness

| Notation | Definition |
|---|---|
| $\theta_j$ | Local relative effectiveness (LRE) of the organization at site *j* |
| $\sqrt{var(\theta_j)}$ | Between-site standard deviation of LRE |
| $Z_{ij}$ | Treatment assignment for individual $i$ at site *j* |
| $Y_{ij}(0)$ | Potential outcome of individual $i$ at site $j$ if assigned to the control condition |
| $Y_{ij}(1)$ | Potential outcome of individual $i$ at site $j$ if assigned to the experimental condition |
| $\delta_{ij}$ | Individual-specific causal effect $Y_{ij}(1) - Y_{ij}(0)$ for individual $i$ at site *j* |
| $\mu_{0j}$ | Average potential outcome of the eligible population of individuals at site $j$ if assigned to the control condition |
| $\mu_{1j}$ | Average potential outcome of the eligible population of individuals at site $j$ if assigned to the experimental condition |
| $\delta_j$ | Site-specific ITT effect $\mu_{1j} - \mu_{0j}$ at site *j* |
| $\mathbf{X}_{ij}$ | Observed pretreatment covariates for individual $i$ at site *j* |
| $\mathbf{U}_{ij}$ | Unobserved pretreatment covariates for individual $i$ at site *j* |
| $\mathbf{\Phi}_{\mathbf{X}j}$ | Parameters defining the distribution of **X** at site $j$, including site means $\mu_{\mathbf{X}j}$ |
| $\mathbf{\Phi}_{\mathbf{U}j}$ | Parameters defining the distribution of **U** at site $j$, including site means $\mu_{\mathbf{U}j}$ |
| $\mathbf{\Phi}_{0j}$ | Parameters characterizing the distribution of $Y(0)$ at site *j* |
| $n_{0j}$ | Sample size of the control group at site *j* |
| $n_{1j}$ | Sample size of the experimental group at site *j* |
| $n_j$ | Sample size at site *j* |
| $\bar{Y}_{0j}$ | Sample mean of the observed outcome of the control group at site *j* |
| $\bar{Y}_{1j}$ | Sample mean of the observed outcome of the experimental group at site *j* |

Table 2. Glossary for Mixed-Effects Model Specification

| Notation | Definition |
|---|---|
| $\sigma_0$ | Within-site standard deviation of the outcome in the control group |
| $\sigma_1$ | Within-site standard deviation of the outcome in the experimental group |
| $\eta_{0j}$ | Site-specific increment to the predicted control group mean outcome |
| $\eta_{0j}^*$ | Empirical Bayes estimator (i.e., the posterior mean) of $\eta_{0j}$ |
| $\omega_{00}$ | The variance of $\eta_{0j}$, which is the between-site variation in the control group mean unexplained by $\mathbf{\Phi}_{\mathbf{X}j}$ |
| $\upsilon_{1j}$ | Site-specific increment to the predicted ITT effect at site $j$ |
| $\upsilon_{1j}^*$ | Empirical Bayes estimator (i.e., the posterior mean) of $\upsilon_{1j}$ |
| $\tau_{11}$ | The variance of $\upsilon_{1j}$, which is the between-site variation in the ITT effect unexplained by $\mathbf{\Phi}_{\mathbf{X}j}$ and $\mathbf{\Phi}_{0j}$ |

Table 3. Site-Level Summary Statistics

| | Site Level | | | | |
|---|---|---|---|---|---|
| Covariate | Mean | SD | P25 | P50 | P75 |
| Number of Observations | 100.0 | 74.9 | 51.0 | 85.5 | 113.5 |
| Treated | 60.7 | 46.0 | 31.8 | 50.5 | 71.3 |
| Control | 39.3 | 29.4 | 20.0 | 33.0 | 45.3 |
| Treatment Share | 60.5% | 5.5 pp. | 57.5% | 60.3% | 63.4% |
| Weekly Earnings in Year 4 | $204 | $34 | $185 | $202 | $225 |
| Treated | $210 | $43 | $185 | $209 | $231 |
| Control | $196 | $40 | $169 | $196 | $227 |
| Female | 39.2% | 18.5 pp. | 24.1% | 39.8% | 52.6% |
| Age | 18.7 | 0.5 | 18.4 | 18.7 | 19.0 |
| 16-17 | 43.7% | 10.8 pp. | 37.3% | 44.5% | 50.6% |
| 18-19 | 30.7% | 6.6 pp. | 26.4% | 30.2% | 36.2% |
| 20-24 | 25.6% | 8.1 pp. | 20.3% | 25.7% | 30.4% |
| Earnings in Prior Year | $2,654 | $971 | $2,110 | $2,580 | $3,103 |
| $0 | 34.9% | 9.6 pp. | 28.4% | 34.2% | 39.6% |
| $1 - $1,000 | 10.6% | 3.4 pp. | 8.5% | 10.2% | 12.8% |
| $1,001 - $5,000 | 26.7% | 6.1 pp. | 22.2% | 26.5% | 30.2% |
| $5,001 - $10,000 | 12.8% | 4.5 pp. | 10.4% | 12.8% | 15.4% |
| Over $10,000 | 6.2% | 3.4 pp. | 4.0% | 6.2% | 8.3% |
| Missing | 8.8% | 4.1 pp. | 5.8% | 8.4% | 10.9% |
| Months Employed in Prior Year | 6.0 | 0.7 | 5.5 | 5.9 | 6.4 |
| 0 - 3 | 17.4% | 5.0 pp. | 14.3% | 17.7% | 20.3% |
| 3 - 9 | 25.3% | 6.0 pp. | 21.9% | 25.8% | 29.1% |
| 9 - 12 | 15.6% | 5.7 pp. | 11.7% | 15.1% | 18.9% |
| Missing | 41.7% | 8.6 pp. | 36.4% | 42.1% | 45.9% |
| Race/Ethnicity | | | | | |
| Black | 42.9% | 29.5 pp. | 13.1% | 43.7% | 67.1% |
| Hispanic | 16.1% | 18.7 pp. | 4.6% | 9.1% | 20.3% |
| White | 32.4% | 23.3 pp. | 14.3% | 26.7% | 49.5% |
| Missing/Other | 8.5% | 12.9 pp. | 2.1% | 4.3% | 10.8% |

*Notes*: pp. stands for percentage points; P25, P50, P75 represent the 25th, 50th, and 75th percentiles, respectively. Age, baseline annual earnings, and number of months employed in the previous year were measured on both categorical and continuous scales. We created missing indicators to capture the mean difference in the outcome between individuals with missing covariate values and those whose covariate values were observed. About nine percent of the participants in the sample did not report previous year's earnings at the baseline; we combined them with those who reported $0 earnings, the latter being the mode of this covariate. In addition, race/ethnicity was either missing or in the "other" category for about nine percent of the participants; we combined them with "Whites" to be the reference category, leaving "Blacks" and "Hispanics" as two distinct minority categories. See section III of Schochet, Burghardt, & McConnell, 2008 for details on the data.

Table 4. Analytic Results of the *2SME* model

| Outcome: Earnings in Year 4 | Step 1 | Step 2 |
|---|---|---|
| EB: Control Group Earnings | | 7.29***<br>(0.81) |
| Treatment | | 15.79**<br>(4.63) |
| EB: Control Group Earnings × Treatment | | -7.42***<br>(1.21) |
| Gender and Race | X | X |
| Age, Baseline Earnings and Employment | X | X |
| Gender and Race × Treatment | | X |
| Age, Baseline Earnings and Employment × Treatment | | X |
| Constant | | 194.24***<br>(3.10) |
| Variance: Random Intercept | 127.28 | 0.86 |
| Variance: Random Slope for Treatment | | 429.57 |
| Covariance: Random Intercept and Random Slope | | 19.20 |
| *N* | 3,929 | 10,001 |

*Notes*: Standard errors in parentheses; ** $p < 0.01$; * $p < 0.001$. "X" indicates the inclusion of the covariates in the model.

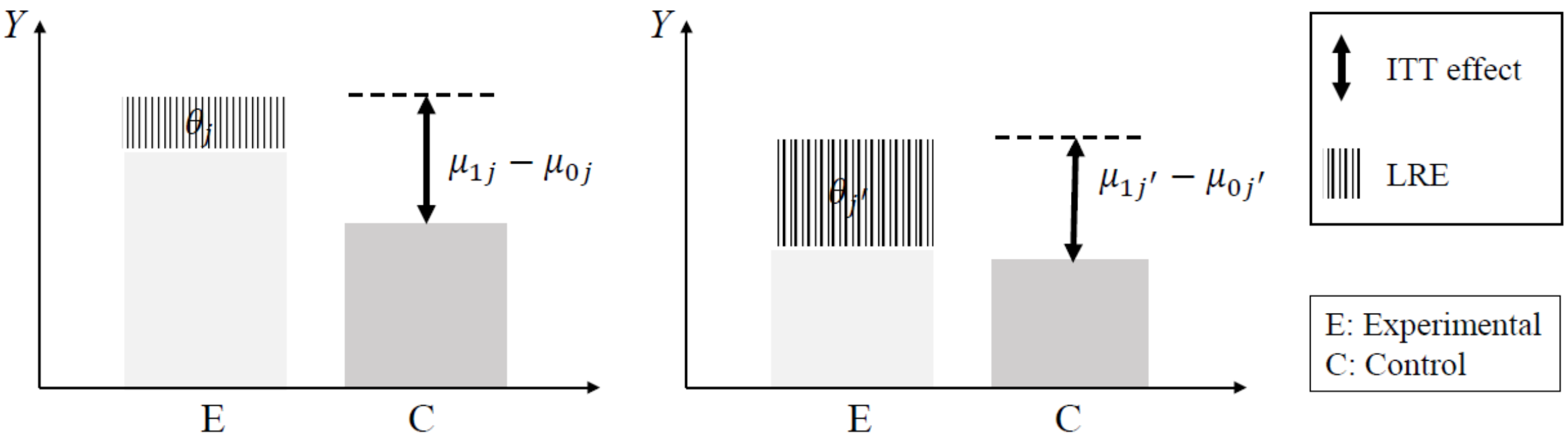

*Notes*: The graph shows that site $j$ does not have the same level of organizational effectiveness as site $j'$ ($\theta_j < \theta_{j'}$), despite having the same ITT effect, because of pre-existing differences in local ecological conditions (e.g., client age composition and local unemployment rate).

Figure 1a. Same ITT Effect and Different Levels of Organizational Effectiveness

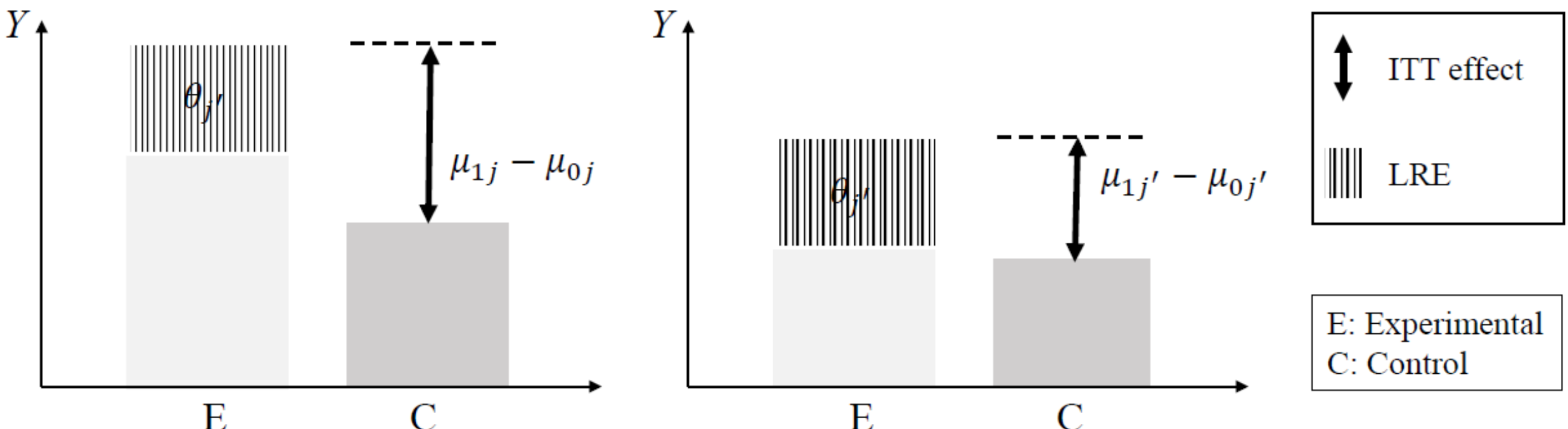

*Notes*: The graph shows that even though sites $j$ and $j'$ have the same level of organizational effectiveness ($\theta_j = \theta_{j'}$), the former has a relatively higher ITT effect because its local ecological conditions are more favorable.

Figure 1b. Different ITT Effects and Same Level of Organizational Effectiveness

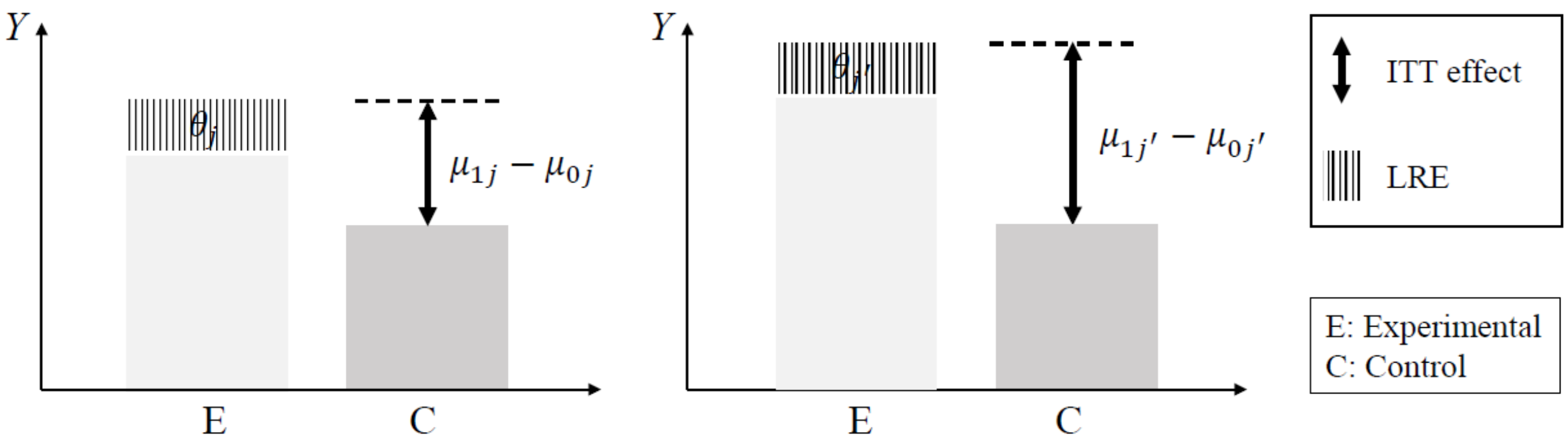

*Notes*: The graph shows that even though sites $j$ and $j'$ have the same level of organizational effectiveness ($\theta_j = \theta_{j'}$), the former has a relatively lower ITT effect because its local ecological conditions are less favorable.

Figure 1c. Different ITT Effects and Same Level of Organizational Effectiveness

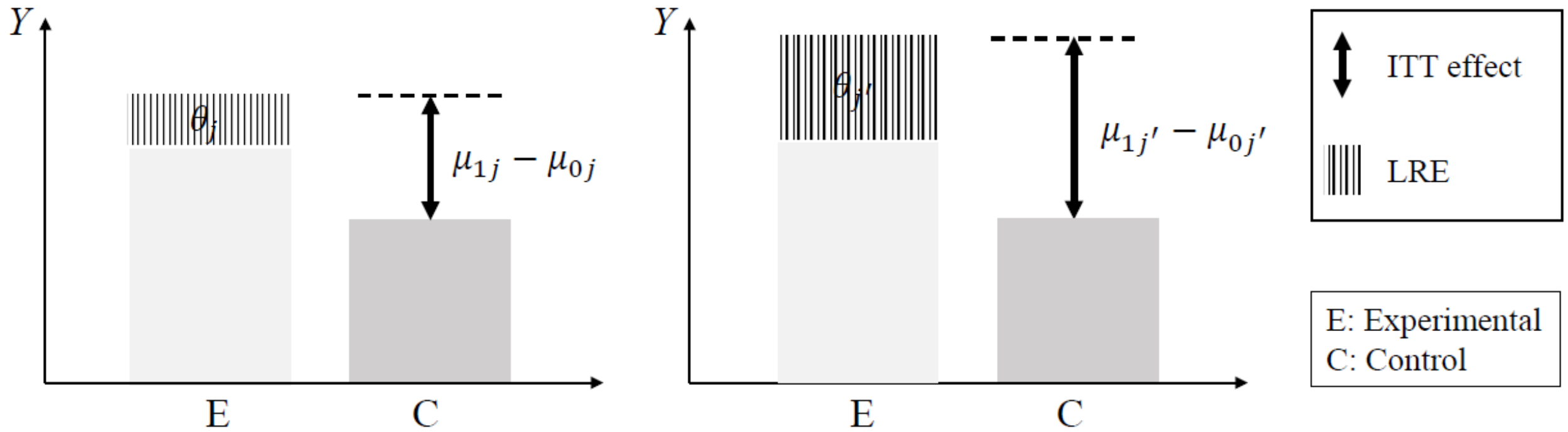

*Note*: The figure shows a case where $\mu_{0j} = \mu_{0j'}$ and $\mu_{1j} - \theta_j = \mu_{1j'} - \theta_{j'}$.

Figure 2. Difference in ITT Equal to Difference in LRE in the Absence of Confounding

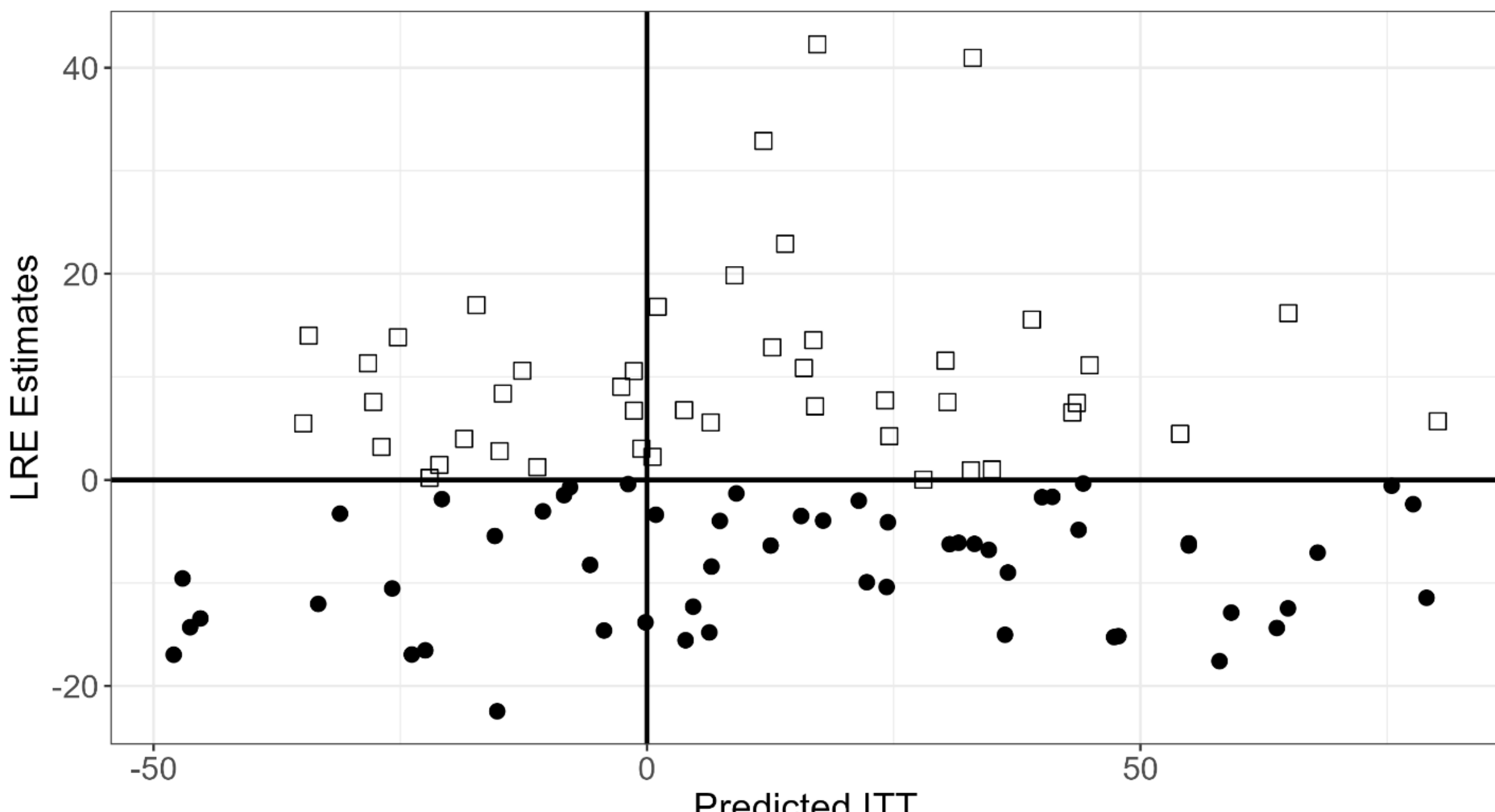

*Notes*: This figure presents a scatterplot of the LRE estimates and the predicted ITT effect estimates of the Job Corps centers, obtained from analyzing the 2SME model; solid dots indicate Job Corps centers with LRE estimates below zero.

Figure 3. Comparing LRE between Sites with Comparable Local Ecological Conditions

**Evaluating Organizational Effectiveness: A New Strategy to Leverage Multisite Randomized Trials for Valid Assessment**

Guanglei Hong[1], Jonah Deutsch[2*], Peter Kress[2*], Jose Eos Trinidad[3*], Zhengyan Xu[4*]

[1] University of Chicago; [2] Mathematica; [3] University of California-Berkeley; [4] University of Pennsylvania

* Names of the co-authors are in alphabetical order.

## Supplementary Material

## I. Simulation Plan

Through a simulation study, we address two empirical questions: (1) How does the proposed procedure compare with alternative strategies when Assumption 1 is mildly violated? (2) How do they compare when Assumption 1 is severely violated?

In the full simulation study, we compare the simulation results across seven candidate strategies for estimating the LRE. They are listed as follows:

*ITT*: a site-by-site OLS analysis of the site-specific ITT effects;
*ITT_adj*: a fixed-effects model adjusting for observed individual-level covariates $\mathbf{X}_{ij}$;
*ME*: a mixed-effects model without adjustment for any covariates;
*ME_adj_X*: a mixed-effects model adjusting for $\mathbf{\Phi}_{\mathbf{X}j}$ only;
*ME_adj_X_Y0*: a mixed-effects model adjusting for $\mathbf{\Phi}_{\mathbf{X}j}$ and $\bar{Y}_{0j}$;
*2SME*: a two-step mixed-effects modeling procedure, which is our proposed strategy;
*ME_adj_X_U*: a mixed-effects model adjusting for $\mathbf{\Phi}_{\mathbf{X}j}$ and $\mathbf{\Phi}_{\mathbf{U}j}$, which is infeasible in practice.

Our primary interest is in (a) comparing *ITT* with *2SME*, (b) comparing *ME_adj_X* with *2SME*, and finally, (c) comparing *2SME* with *ME_adj_X_U*. Hence the manuscript includes graphs that compare these four candidate strategies. The supplementary material provides additional results.

### I.a. Data Generation

We generate simulation data that resemble the NJCS data with 100 sites ( $J = 100$ ) and 100 individuals at each site on average. However, the design is imbalanced—sample size varies across the sites. The sample size at a site is randomly drawn from a uniform distribution between 30 and 170.

A larger sample size per site is expected to remove a greater amount of the finite sample bias. Hence, we generate another set of simulation data with 100 sites and 700 individuals at each site on average. The sample size at a site is again randomly drawn from a uniform distribution between 400 and 1000.

We further examine an extreme case with only 20 individuals per site on average or with only 30 sites ($J = 30$) in total.

***Generating observed and unobserved covariates.*** The data generation starts with site-level and individual-level observed and unobserved pretreatment covariates. Let $\mathbf{X} = (X_1 \quad X_2)$ and $\mathbf{U} = (U_1 \quad U_2)$ where $X_1$, $X_2$, $U_1$, and $U_2$ are all independent. We generate $\mu_{\mathbf{X}j}$ and $\mu_{\mathbf{U}j}$ that each follow a bivariate normal distribution over all the sites. Specifically, the distributions of $\mu_{X_1j}$, $\mu_{X_2j}$, $\mu_{U_1j}$, and $\mu_{U_2j}$ are each $N(0, 0.1)$. Within site $j$, $\mathbf{X}_{ij}$ and $\mathbf{U}_{ij}$ each follow a bivariate normal distribution with their means equal to $\mu_{\mathbf{X}j}$ and $\mu_{\mathbf{U}j}$, respectively:
$X_{1ij} \sim N\left(\mu_{X_1j}, 1\right)$; $X_{2ij} \sim N\left(\mu_{X_2j}, 1\right)$; $U_{1ij} \sim N\left(\mu_{U_1j}, 1\right)$; $U_{2ij} \sim N\left(\mu_{U_2j}, 1\right)$.

***Generating potential outcomes.*** Corresponding to the two empirical research questions, we create two different scenarios in data generation. In Scenario 1, $\mathbf{U}_{ij}$ predicts both $Y_{ij}(0)$ and $Y_{ij}(1) - Y_{ij}(0)$, which represents a relatively minor violation of Assumption 1. In scenario 2, only $U_{1ij}$ predicts both $Y_{ij}(0)$ and $Y_{ij}(1) - Y_{ij}(0)$; in contrast, $U_{2ij}$ predicts $Y_{ij}(1) - Y_{ij}(0)$ but not $Y_{ij}(0)$, which represents a severe violation of Assumption 1.

Under scenario 1, the potential outcome $Y_{ij}(0)$ (weekly earnings four years after the randomization) is a function of $\mathbf{X}_{ij}$, $\mathbf{U}_{ij}$, $\mu_{\mathbf{X}j}$, and $\mu_{\mathbf{U}j}$ plus a normal random error. We may allow for negative $Y_{ij}(0)$.

$$Y_{ij}(0) = 197 + 120X_{1ij} - 100X_{2ij} + 60U_{1ij} - 50U_{2ij} + 20\mu_{X_1j} - 30\mu_{X_2j} + 20\mu_{U_1j} - 20\mu_{U_2j} + \varepsilon_{(y)ij}, \qquad \varepsilon_{(y)ij} \sim N(0, 3{,}000).$$

The average within-site standard deviation of $Y_{ij}(0)$, which is expected to be equal to $\sigma = \sqrt{120^2 + (-100)^2 + 60^2 + (-50)^2 + 3{,}000} = \sqrt{33{,}500} \approx 183$, is used as the scaling unit for calculating the effect size of the treatment effect on the outcome.

Under scenario 2, the potential outcome $Y_{ij}(0)$ (weekly earnings four years after the randomization) is a function of $\mathbf{X}_{ij}$, $U_{1ij}$, $\mu_{\mathbf{X}j}$, and $\mu_{U_1j}$ plus a normal random error. Again we may allow for negative $Y_{ij}(0)$.

$$Y_{ij}(0) = 197 + 120X_{1ij} - 100X_{2ij} + 78U_{1ij} + 20\mu_{X_1j} - 30\mu_{X_2j} + 28\mu_{U_1j} + \varepsilon_{(y)ij}, \quad \varepsilon_{(y)ij} \sim N(0, 3{,}000).$$

The average within-site standard deviation of $Y_{ij}(0)$, which is expected to be equal to $\sigma = \sqrt{120^2 + (-100)^2 + 78^2 + 3{,}000} = \sqrt{33{,}484} \approx 183$, is used as the scaling unit for calculating the effect size of the treatment effect on the outcome.

The ITT effect of attending the Job Corps center at site $j$, denoted as $\delta_j$, is a function of $\mu_{\mathbf{X}j}$ and $\mu_{\mathbf{U}j}$ plus $\theta_j$. Here $\theta_j$ represents the relative effectiveness of the center at site $j$.

$$\delta_j = 13 - 70\mu_{X_1j} + 70\mu_{X_2j} - 80\mu_{U_1j} + 90\mu_{U_2j} + \theta_j,$$

$$\theta_j \sim N\left(0, \psi_{\theta_j}^2\right).$$

The average ITT effect is expected to be $13 = 0.07\sigma$ where 0.07 is the effect size. Weiss et al's (2017) re-analysis of data from 16 large-scale multisite randomized trials has shown that the between-site standard deviation of the ITT effect may vary from $0\sigma$ to $0.35\sigma$. In this simulation study, we let the between-site standard deviation of $\theta_j$, denoted as $\psi_{\theta_j}$, take a range of values from $0\sigma$ to $0.35\sigma$. Within site $j$, the individual-specific causal effect $\delta_{ij}$ is a function of $\mathbf{X}_{ij}$ and $\mathbf{U}_{ij}$ plus a normal random error; the mean of $\delta_{ij}$ is expected to be equal to $\delta_j$:

$$\delta_{ij} \sim N\left(\delta_j + 4X_{1ij} + 2.5X_{2ij} - 2U_{1ij} - 1.5U_{2ij}, 4\right).$$

The within-site standard deviation of $\delta_{ij}$ is approximately equal to 6. Finally, the potential outcome $Y_{ij}(1)$ is equal to $Y_{ij}(0) + \delta_{ij}$.

***Generating the treatment assignment indicator.*** For individual $i$ at site $j$, we generate the treatment assignment indicator $Z_{ij} \sim Bernoulli(0.5)$. This data generation mechansim satisfies Assumption 2.

**I.b. Evaluation Criteria**

We aggregate the results including bias, efficiency (empirical variance), and MSE for the site-specific LRE estimates over all the sites. The naïve results are obtained by conducting an ITT analysis site by site and using the site-specific ITT effect obtained from the sample as an estimator of $\theta_j$. We denote the estimated site-specific ITT effect with $\hat{\delta}_j$ for site $j$. The results of every alternative strategy, denoted as $\hat{\theta}_j$ for site $j$, are to be compared with the naïve results through computing the bias, empirical variance, and RMSE.

We further use the severe classification error (SCE) rate and the moderate classification error (MCE) rate as additional evaluation criteria. To do so, we rank the 100 sites according to their estimated values of $\hat{\delta}_j$ or $\hat{\theta}_j$. The ranked sites are then divided into three tiers—30% at the high level, 40% at the medium level, and 30% at the low level. These are to be compared with the tier membership based on the true value of LRE. A severe classification error occurs when a site in the top tier according to the true value of its LRE is mistakenly placed in the bottom tier or vice versa. A moderate classification error occurs when a site in the top or bottom tier is mistakenly placed in the middle tier or vice versa.

***Bias.*** For the naïve results, let $Bias_{j.naive}$ denote the bias in the site-specific ITT effect estimator $\hat{\delta}_j$ for site $j$ that has been averaged over all the 500 simulations. We use $Bias_{j.naive} = \hat{E}\left[\hat{\delta}_j\right] - \theta_j$ to represent this average for site $j$.

$$Mean\ of\ Bias_{naive} = E\left[Bias_{j.naive}\right] = \frac{1}{J}\sum_{j=1}^{J}\left(\hat{E}\left[\hat{\delta}_j\right] - \theta_j\right);$$

$$Standard\ Deviation\ of\ Bias_{naive} = \sqrt{\frac{1}{J-1}\sum_{j=1}^{J}\left\{Bias_{j.naive} - E\left[Bias_{j.naive}\right]\right\}^2}.$$

For the results from each one of the adjustment strategies, we replace $\hat{\delta}_j$ with $\hat{\theta}_j$ in computing the $Bias_{j.adjusted}$, ${Bias_{j.adjusted}}^2$, and the mean and standard deviation of $Bias_{adjusted}$. In reporting the simulation results, we represent the mean and standard deviation of $Bias_{naive}$ and $Bias_{adjusted}$ each as a fraction of $\sigma$.

***Efficiency.*** The efficiency of the naive estimator $\hat{\delta}_j$ for site $j$ is obtained by computing the empirical variance over all the simulations:

$$\widehat{var}(\hat{\delta}_j) = \hat{E}\left[\left(\hat{\delta}_j - \hat{E}[\hat{\delta}_j]\right)^2\right].$$

This is then averaged over all the sites to obtain $Average\ Empirical\ Variance_{naive}$:

$$Average\ Empirical\ Variance_{naive} = \frac{1}{J}\sum_{j=1}^{J}\widehat{var}(\hat{\delta}_j) = \frac{1}{J}\sum_{j=1}^{J}\hat{E}\left[\left(\hat{\delta}_j - \hat{E}[\hat{\delta}_j]\right)^2\right].$$

Again, for the results from each one of the adjustment strategies, we replace $\hat{\delta}_j$ with $\hat{\theta}_j$ in computing $var(\hat{\theta}_j)$ and $Average\ Empirical\ Variance_{adjusted}$. In reporting the simulation results, we express $Average\ Empirical\ Variance_{naive}$ and $Average\ Empirical\ Variance_{adjusted}$ each as a fraction of $\sigma^2$. The variance ratio is computed as follows:

$$Variance\ Ratio = \frac{Average\ Empirical\ Variance_{adjusted}}{Average\ Empirical\ Variance_{naive}}.$$

Among the competing adjustment strategies, we will identify the one that displays the lowest variance ratio.

***Root mean square error.*** The RMSE of the naive estimator $\hat{\delta}_j$ for site $j$ is

$$RMSE(\hat{\delta}_j) = \sqrt{\hat{E}\left[\left(\hat{\delta}_j - \theta_j\right)^2\right]};$$

$$Average\ RMSE_{naive} = \frac{1}{J}\sum_{j=1}^{J}\sqrt{\hat{E}\left[\left(\hat{\delta}_j - \theta_j\right)^2\right]}.$$

Because $\hat{E}\left[\left(\hat{\delta}_j - \theta_j\right)^2\right] = \hat{E}\left[\left(\hat{\delta}_j - \hat{E}[\hat{\delta}_j] + \hat{E}[\hat{\delta}_j] - \theta_j\right)^2\right] = \hat{E}\left[\left(\hat{\delta}_j - \hat{E}[\hat{\delta}_j]\right)^2\right] + \left(\hat{E}[\hat{\delta}_j] - \theta_j\right)^2$, $Average\ RMSE_{naive} = \frac{1}{J}\sum_{j=1}^{J}\sqrt{\widehat{var}(\hat{\delta}_j) + {Bias_{j.naive}}^2}$. For the results from each one of the adjustment strategies, we replace we replace $\hat{\delta}_j$ with $\hat{\theta}_j$ in computing $RMSE(\hat{\theta}_j)$ and $Average\ RMSE_{adjusted}$. In reporting the simulation results, we express $Average\ RMSE_{naive}$ and $Average\ RMSE_{adjusted}$ each as a fraction of $\sigma$. The Average RMSE Percent Reduction is computed as follows.

$$Average\ RMSE\ Percent\ Reduction = 1 - \frac{Average\ RMSE_{adjusted}}{Average\ RMSE_{naive}}$$

Among the competing adjustment strategies, we will identify the one that displays the highest mean RMSE percent reduction.

***Severe classification error rate.*** Let $L_j = l$ denote the true level of organizational effectiveness at site $j$, where $l = 1, 2, 3$ correspond to the low-, medium-, and high-level of LRE. After conducting the naïve analysis in each simulation, we sort all the sites in an ascending order according to the value of $\hat{\delta}_j$. Let $\hat{L}_{j.naive} = 1$ if $\hat{\delta}_j$ is in the bottom 30%; $\hat{L}_{j.naive} = 2$ if $\hat{\delta}_j$ is in the middle 40%; and $\hat{L}_{j.naive} = 3$ if $\hat{\delta}_j$ is in the top 30% of the distribution. A severe classification error occurs when a Job Corps center in the top tier according to its true level of organizational effectiveness is mistakenly placed in the bottom tier or vice versa. Let $SCE_{j.naive}$ be a dummy indicator that takes value 1 if $\left(\hat{L}_{j.naive} - L_j\right)^2 = 4$ and 0 otherwise. We obtain the SCE rate for site $j$ that is averaged over all the 500 simulations and use $\hat{E}\left[SCE_{j.naive}\right]$ to represent this average. The mean SCE rate over all the sites is:

$$Mean\ SCE\ Rate_{naive} = \frac{1}{J}\sum_{j=1}^{J} \hat{E}\left[SCE_{j.naive}\right].$$

***Moderate classification error rate.*** A moderate classification error occurs when a center in the top or bottom tier is mistakenly placed in the middle tier or vice versa. Let $MCE_{j.naive}$ be a dummy indicator that takes value 1 if $\left(\hat{L}_{j.naive} - L_j\right)^2 = 1$ and 0 otherwise. We obtain the MCE rate for site $j$ that is averaged over all the 500 simulations and use $\hat{E}\left[MCE_{j.naive}\right]$ to represent this average. The mean MCE rate over all the sites is:

$$Mean\ MCE\ Rate_{naive} = \frac{1}{J}\sum_{j=1}^{J} \hat{E}\left[MCE_{j.naive}\right].$$

## II. Supplementary Simulation Results

To address research question (1), we set up a scenario in which the unobserved covariates $\mathbf{U}_{ij}$ are no less important than the observed covariates $\mathbf{X}_{ij}$ in predicting $\delta_{ij}$ as well as $Y_{ij}(0)$. Research question (2) is addressed through a second scenario where only $U_{1ij}$ predicts both $\delta_{ij}$ and $Y_{ij}(0)$; in contrast, $U_{2ij}$ predicts $\delta_{ij}$ but not $Y_{ij}(0)$ and thus Assumption 1 is clearly violated. In practice, when sites differ in not just the mean but also the variance or other higher dimensions of the distribution of $\mathbf{X}_{ij}$ or $Y_{ij}(0)$, omitting these higher dimensions or misspecifying the functional forms in statistical adjustment can be viewed as a special case of the second scenario.

### II.a. Scenario 1 with a Minor Violation of Assumption 1

For scenario 1 in which the violation of Assumption 1 is relatively minor, we display the simulation results when $J = 100$. For each evaluation criterion, we compare the results when $n_j = 100$ while the number of sites varies ($J = 100$ vs. $J = 30$); we also compare with the results when $n_j = 20$ or $n_j = 700$ while $J = 100$. The simulation results are displayed across a plausible range of the between-site standard deviation of the LRE. See Figures S1-S6.

### II.b. Scenario 2 with a Severe Violation of Assumption 1

We display the simulation results for scenario 2 in which Assumption 1 is severely violated. Again, for each evaluation criterion, we compare the results when $n_j = 100$ while the number of sites varies ($J = 100$ vs. $J = 30$); we also compare with the results when $n_j = 20$ or $n_j = 700$ while $J = 100$. See Figures S7-S12.

### II.c. Summary of Simulation Results

***Bias***. Figure S1 shows the distribution of bias over the $J$ sites in scenario 1 when the between-site standard deviation of LRE is set to be 0.1. Among the four candidate strategies, the site-specific ITT effect obtained from the fixed-effects *ITT* model has the widest distribution of bias; the mixed-effects model with adjustment for $\mathbf{\Phi}_{\mathbf{X}j}$ (*ME_adj_X*) removes a considerable amount of bias; and the mixed-effects model that infeasibly adjusts for both $\mathbf{\Phi}_{\mathbf{X}j}$ and $\mathbf{\Phi}_{\mathbf{U}j}$ (*ME_adj_X_U*) tends to be least biased.

Impressively, the proposed two-step procedure (*2SME*) removes nearly as much bias as the infeasible model. Figure S2 shows the same pattern as above when the between-site standard deviation of LRE increases from 0 to 0.35 in this scenario. Because a mixed-effects model uses the shrinkage estimator, even the infeasible *ME_adj_X_U* model contains some bias except when the sample size per site goes to infinity or when the LRE is constant across all the sites. As the sample size per site increases, the average magnitude of bias decreases for the *2SME* estimator as well as the infeasible *ME_adj_X_U* estimator. The same pattern does not hold for the estimators obtained from the other two strategies.

Assumption 1 is severely violated in scenario 2. As one may anticipate, the *2SME* model is less successful in bias removal than the infeasible *ME_adj_X_U* model but more successful than the *ME_adj_X* model and the *ITT* model. The *2SME* estimator no longer displays the statistical property of consistency in this scenario. Only the infeasible *ME_adj_X_U* estimator remains consistent. See supplementary materials for details.

***Efficiency***. Consistent with the general anticipation, as the sample size per site increases, the average empirical variance decreases for all the models under consideration. Also as anticipated, when employing any of the three mixed-effects models, an increase in the between-site standard deviation of LRE leads to an increase in the variance of the LRE estimator and thus a reduction in efficiency (Appendix C provides a rationale for this pattern). Although the performance of the *ITT* model is unaffected, it displays the least efficiency within the range of the between-site standard deviation of LRE as shown in Figure S3. In scenario 1, the proposed *2SME* model tends to produce more efficient results than all the comparison models including the infeasible *ME_adj_X_U* model. We reason that this is because $\mathbf{\Phi}_{\mathbf{U}j}$ is multivariate and uses more degrees of freedom than $\mathbf{\Phi}_{0j}$ in estimation. Yet under scenario 2 in which Assumption 1 is severely violated, the *2SME* estimator no longer outperforms the infeasible *ME_adj_X_U* estimator in efficiency although it still generally outperforms the other two candidate strategies.

***RMSE***. Figure S4 shows that the average estimation accuracy increases with the sample size per site for all the models under consideration in scenario 1. The proposed *2SME* estimator shows a lower value of average RMSE and thus higher overall accuracy than nearly all the comparison models especially when the sample size per site is relatively small. However, under scenario 2, the *2SME* estimator becomes less accurate than the infeasible *ME_adj_X_U* estimator but is still more accurate than the *ITT* estimator and the *ME_adj_X* estimator.

***SCE and MCE rates***. Figures S5 and S6 show that, as the between-site standard deviation of LRE increases, LRE values become increasingly distinguishable between the sites; and hence the SCE and MCE rates decrease. Under scenario 1 in which the violation of Assumption 1 is minor, the *2SME* estimator tends to have SCE and MCE rates as low as or even lower than the infeasible *ME_adj_X_U* estimator. As the sample size per site increases, the SCE and MCE rates further decrease for the *2SME* model and the infeasible *ME_adj_X_U* model due to the decrease in bias; this pattern does not hold for the *ITT* model and the *ME_adj_X* model. For example, with $n_j = 100$ and the between-site standard deviation of LRE at or above 0.1 in standardized units, the *2SME* model would severely misclassify only about 2% of the organizations; under the same conditions, the *ITT* model and the *ME_adj_X* model would severely misclassify about six times as many organizations. With $n_j = 700$, the *2SME* model would have zero SCE rate while the *ITT* model and the *ME_adj_X* model would still severely misclassify close to 12% of the organizations. The MCE rates are highly sensitive to estimation uncertainty and thus tend to be considerably higher than the SCE rates. Under scenario 2 in which Assumption 1 is severely violated, as anticipated, the SCE and MCE rates from the *2SME* model are higher than the infeasible *ME_adj_X_U* model but nonetheless generally lower than the ITT model and the *ME_adj_X* model.

In summary, the site-specific ITT effect obtained from the fixed-effects *ITT* model has the widest distribution of bias; the mixed-effects model with adjustment for $\mathbf{\Phi}_{\mathbf{X}j}$ (*ME_adj_X*) removes a considerable amount of bias; and the mixed-effects model that infeasibly adjusts for both $\mathbf{\Phi}_{\mathbf{X}j}$ and $\mathbf{\Phi}_{\mathbf{U}j}$ (*ME_adj_X_U*) tends to be least biased under both scenarios. Impressively, the proposed two-step procedure (*2SME*) removes as much bias as the infeasible model. Even when the sample size per site is relatively small, the 2SME estimator still shows higher efficiency, lower RMSE, and lower SCE and MCE rates than the infeasible model. This is likely because, by summarizing the mulviariate information of **U** in the control group mean outcome, the *2SME* strategy consumes fewer degrees of freedom than the infeasible model.

### ***Distribution of Bias***

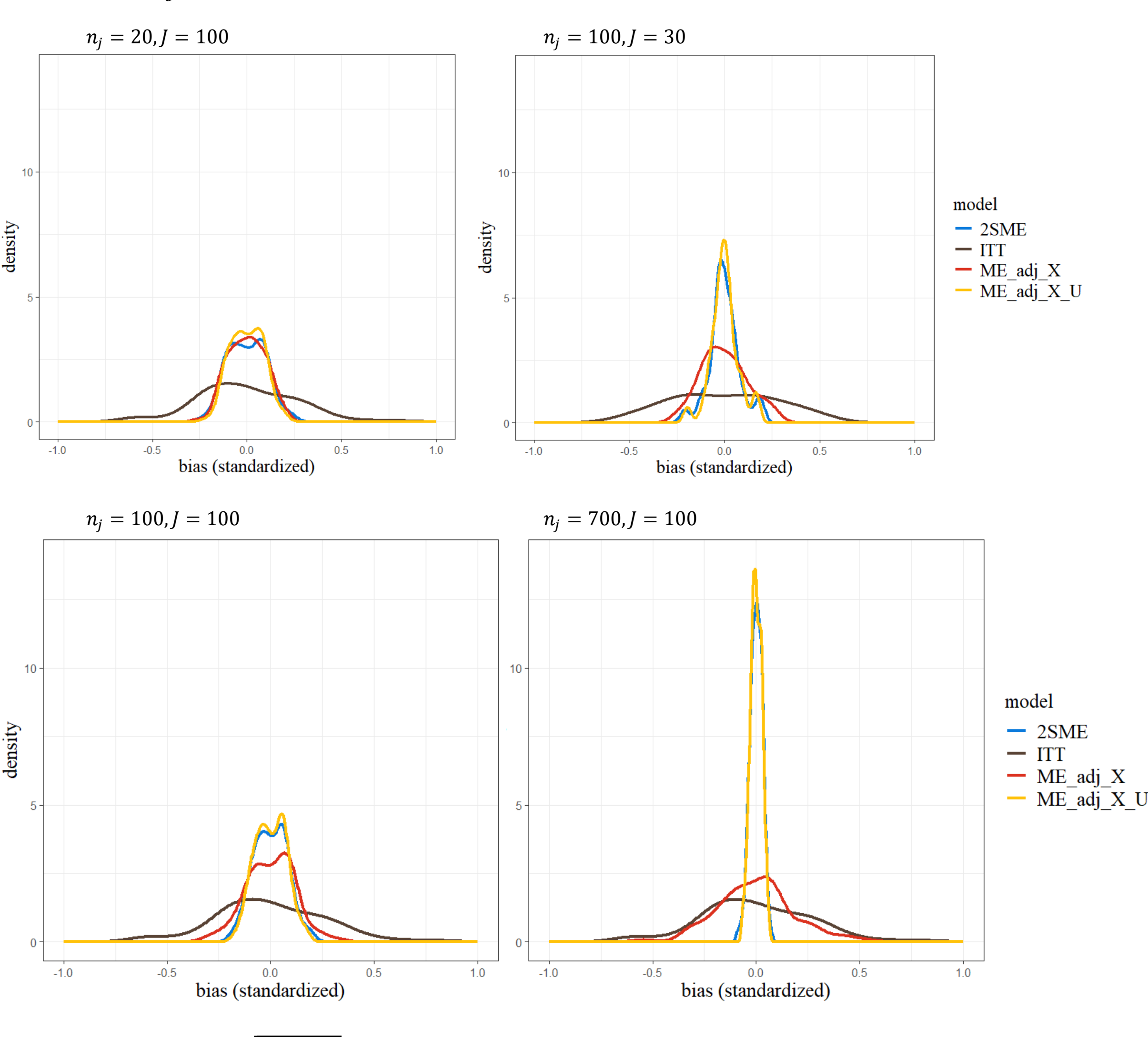

Note: The standardized $\sqrt{var(\theta_j)} = 0.1$.

Figure S1. Between-Model Comparison of the Distribution of Bias in LRE Estimation Under Scenario 1

### ***Standard Deviation of Bias***

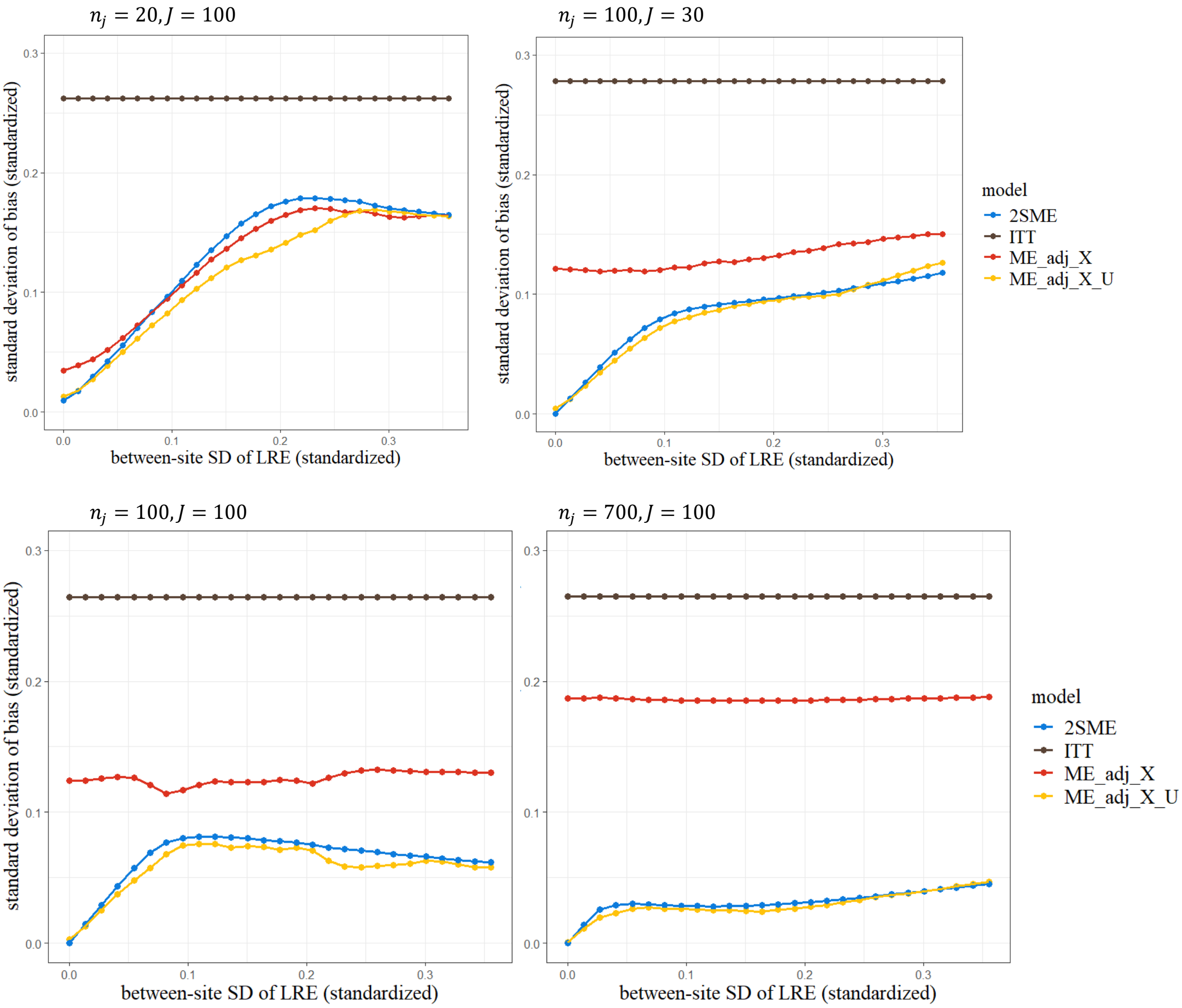

Figure S2. Between-Model Comparison of the Standard Deviation of Bias in LRE Estimation Under Scenario 1

### *Efficiency*

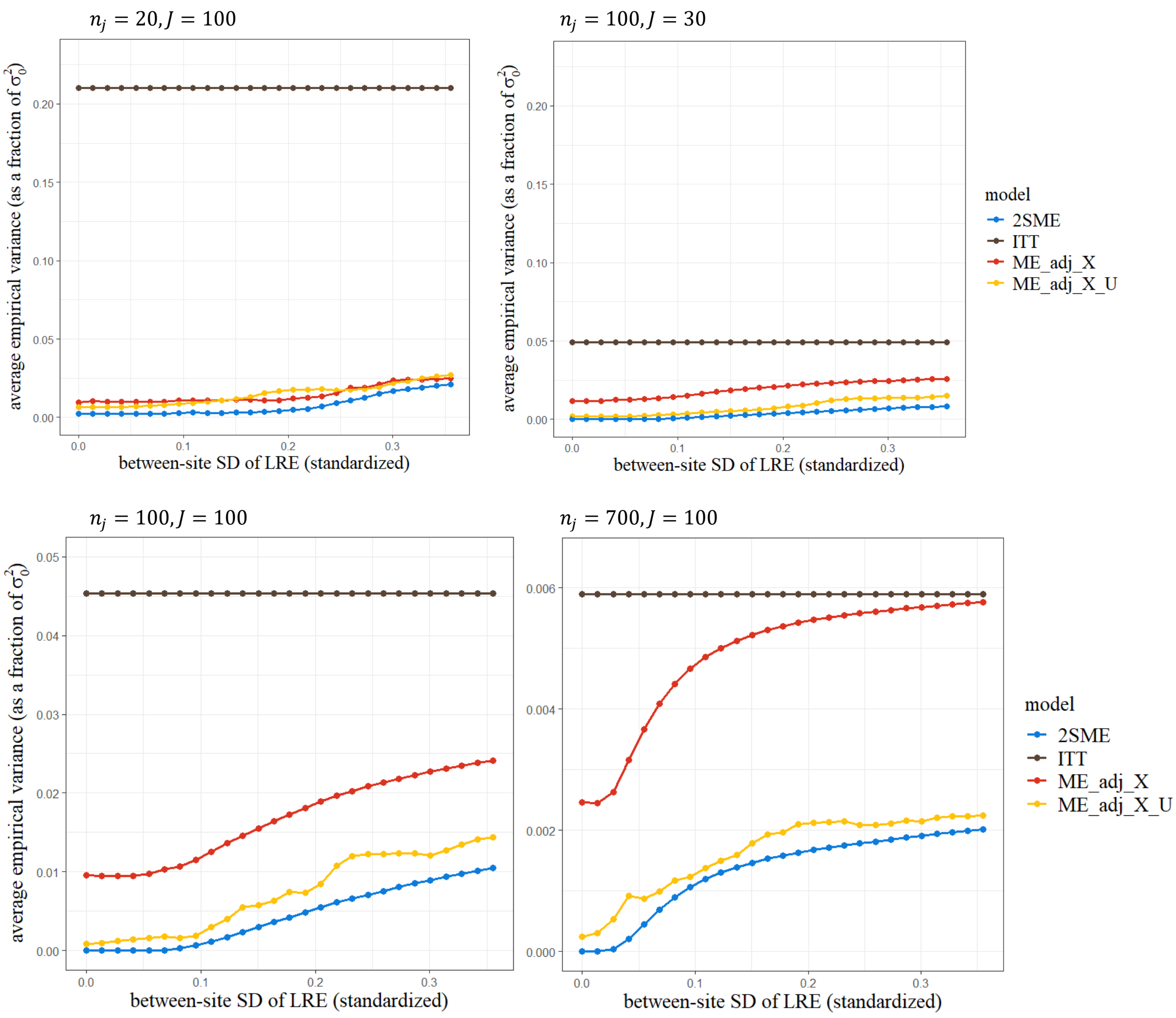

Note: The vertical scales differ between the three sample sizes per site.

Figure S3. Between-Model Comparison of Average Efficiency in LRE Estimation Under Scenario 1

## *RMSE*

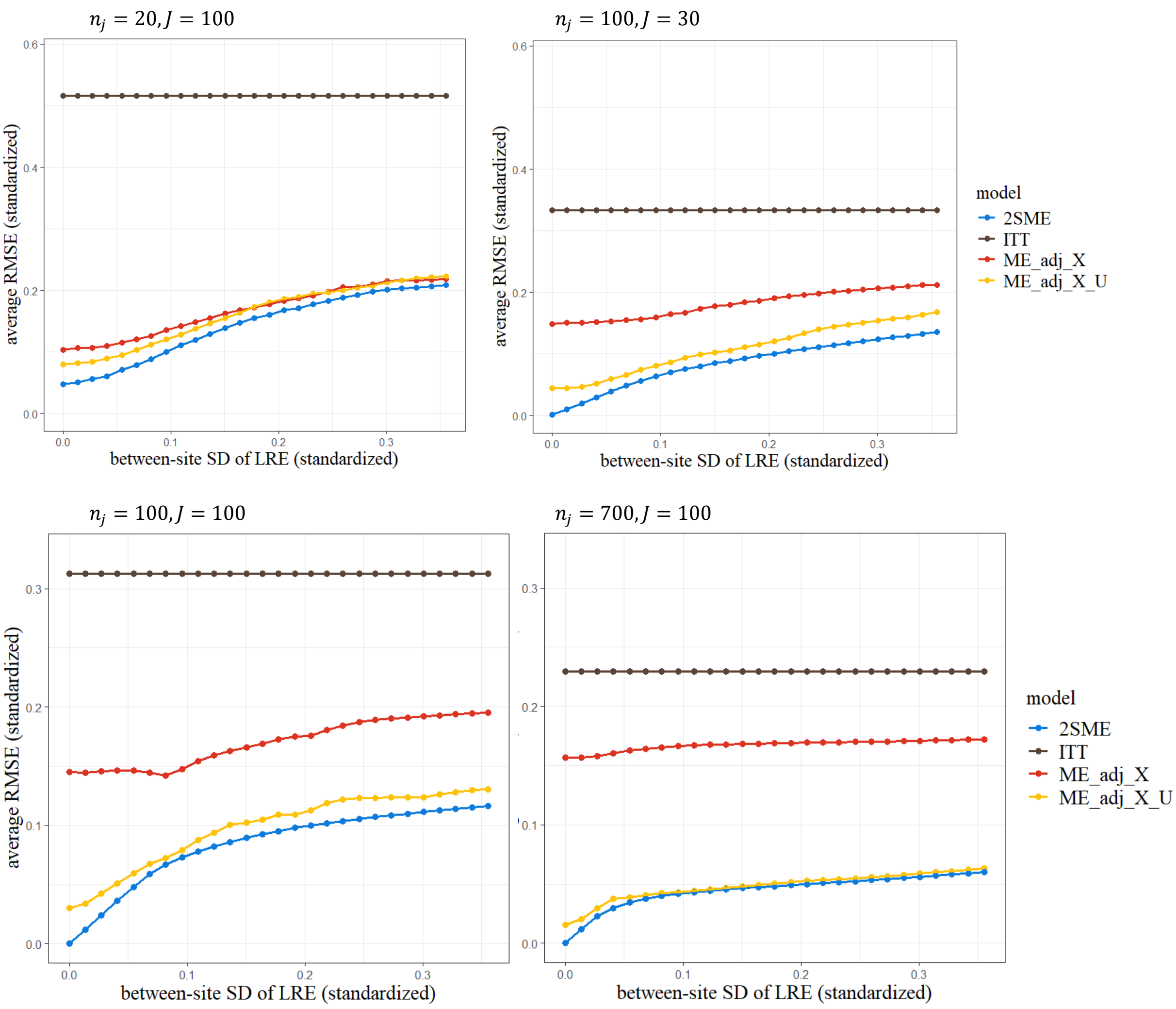

Figure S4. Between-Model Comparison of Average RMSE in LRE Estimation Under Scenario 1

## ***SCE Rate***

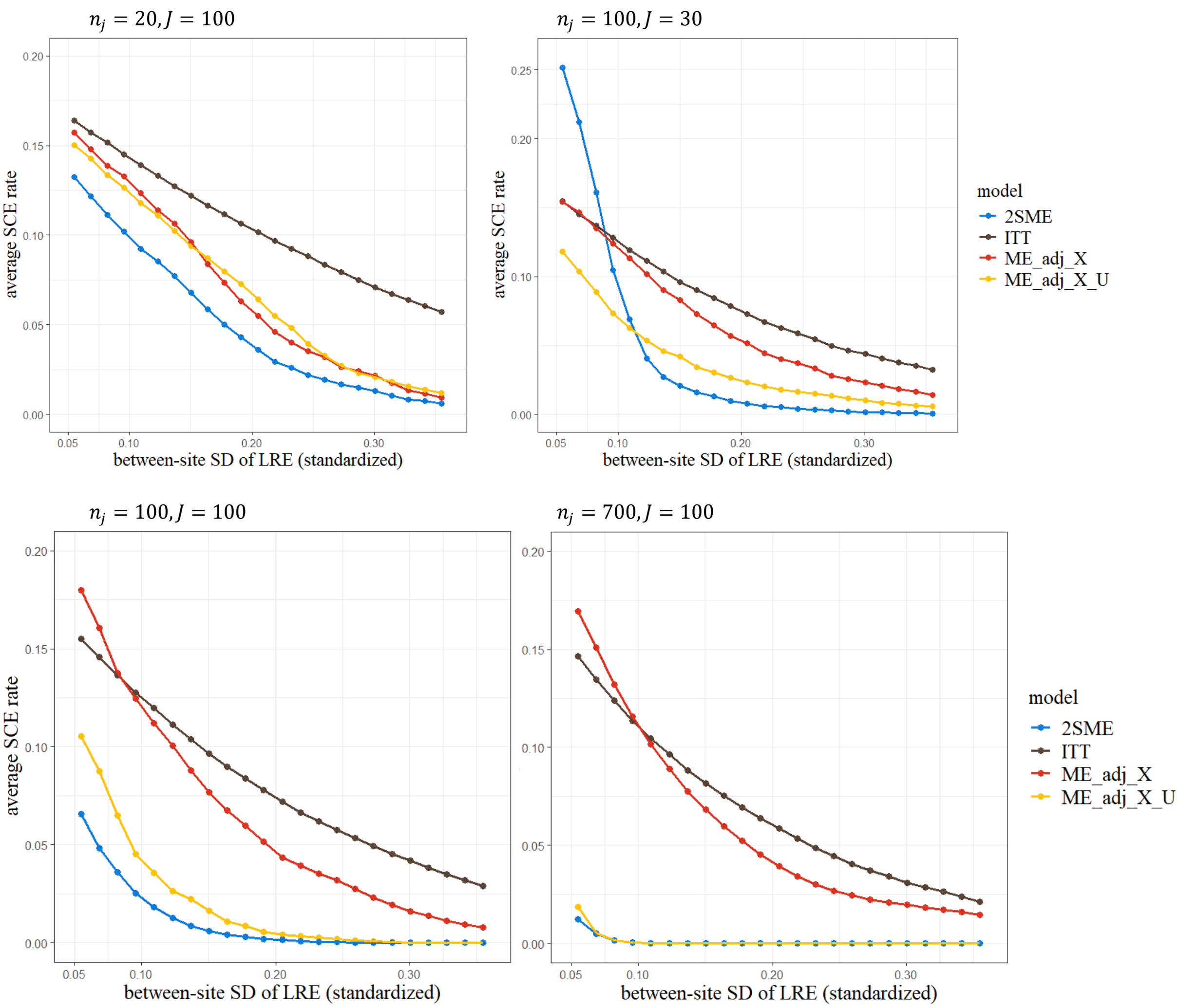

Figure S5. Between-Model Comparison of SCE Rate in LRE Estimation Under Scenario 1

## ***MCE Rate***

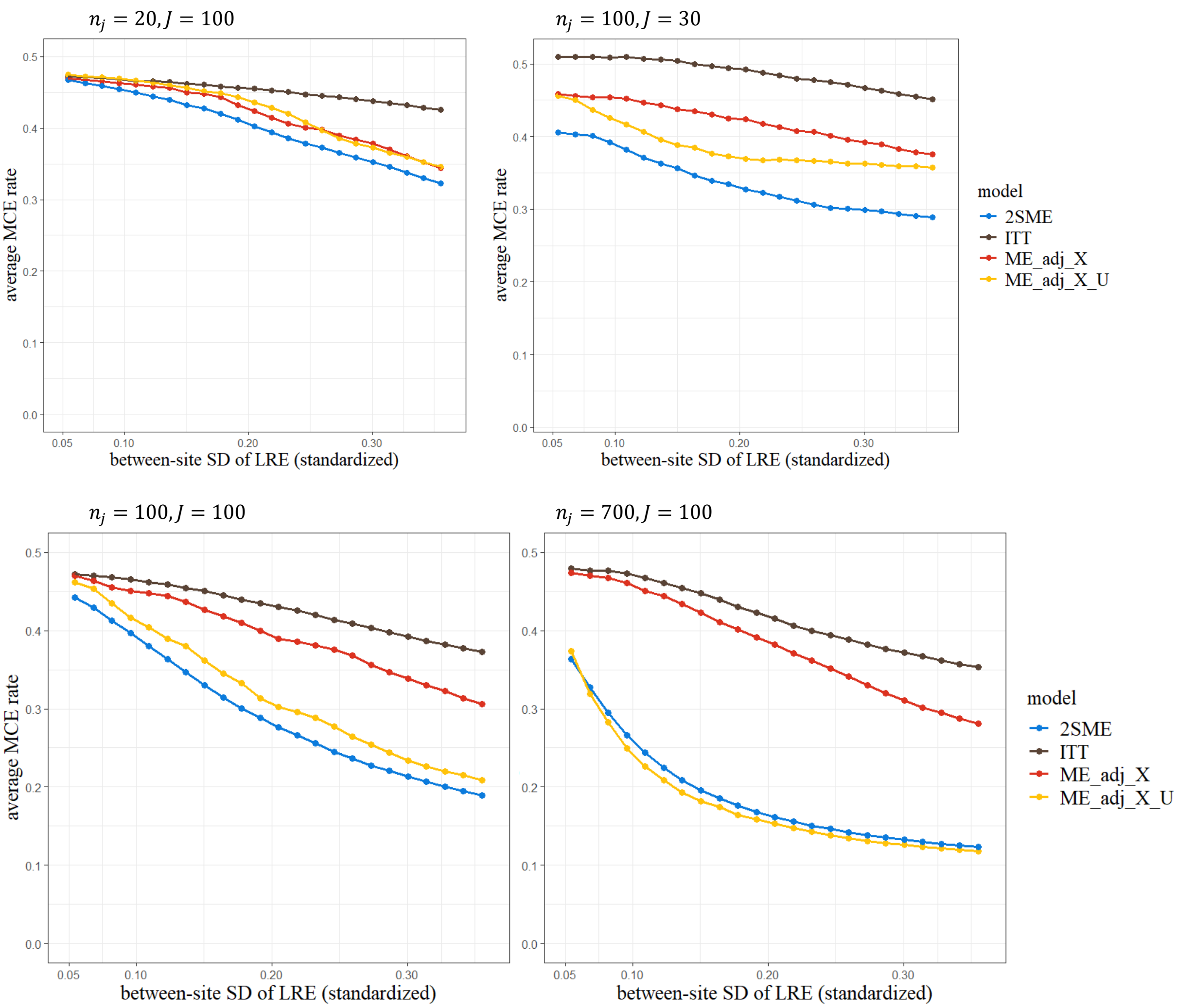

Figure S6. Between-Model Comparison of MCE Rate in LRE Estimation Under Scenario 1

***Distribution of Bias***

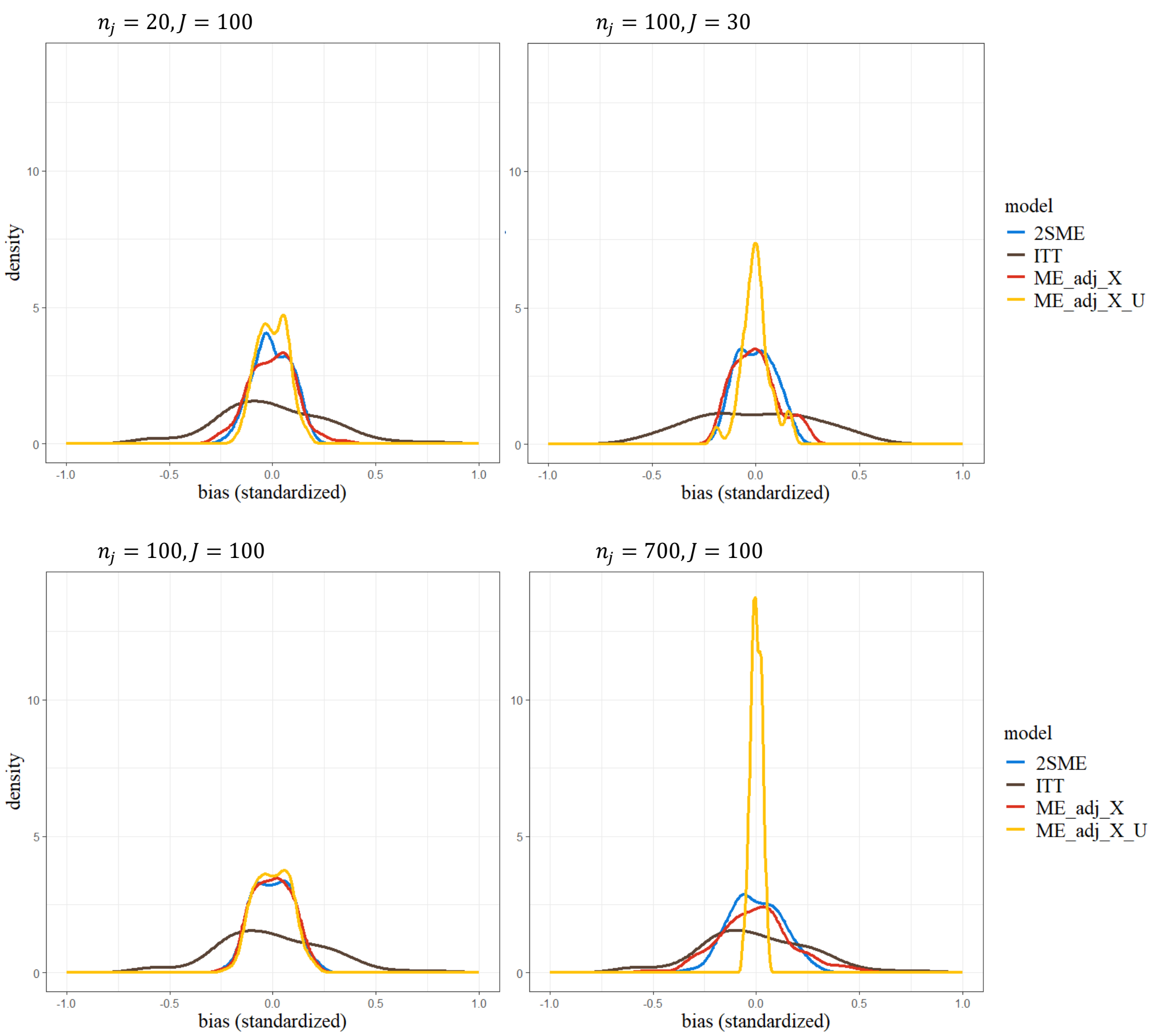

Note: The standardized $\sqrt{var(\theta_j)} = 0.1$.

Figure S7. Between-Model Comparison of the Distribution of Bias in LRE Estimation Under Scenario 2

### ***Standard Deviation of Bias***

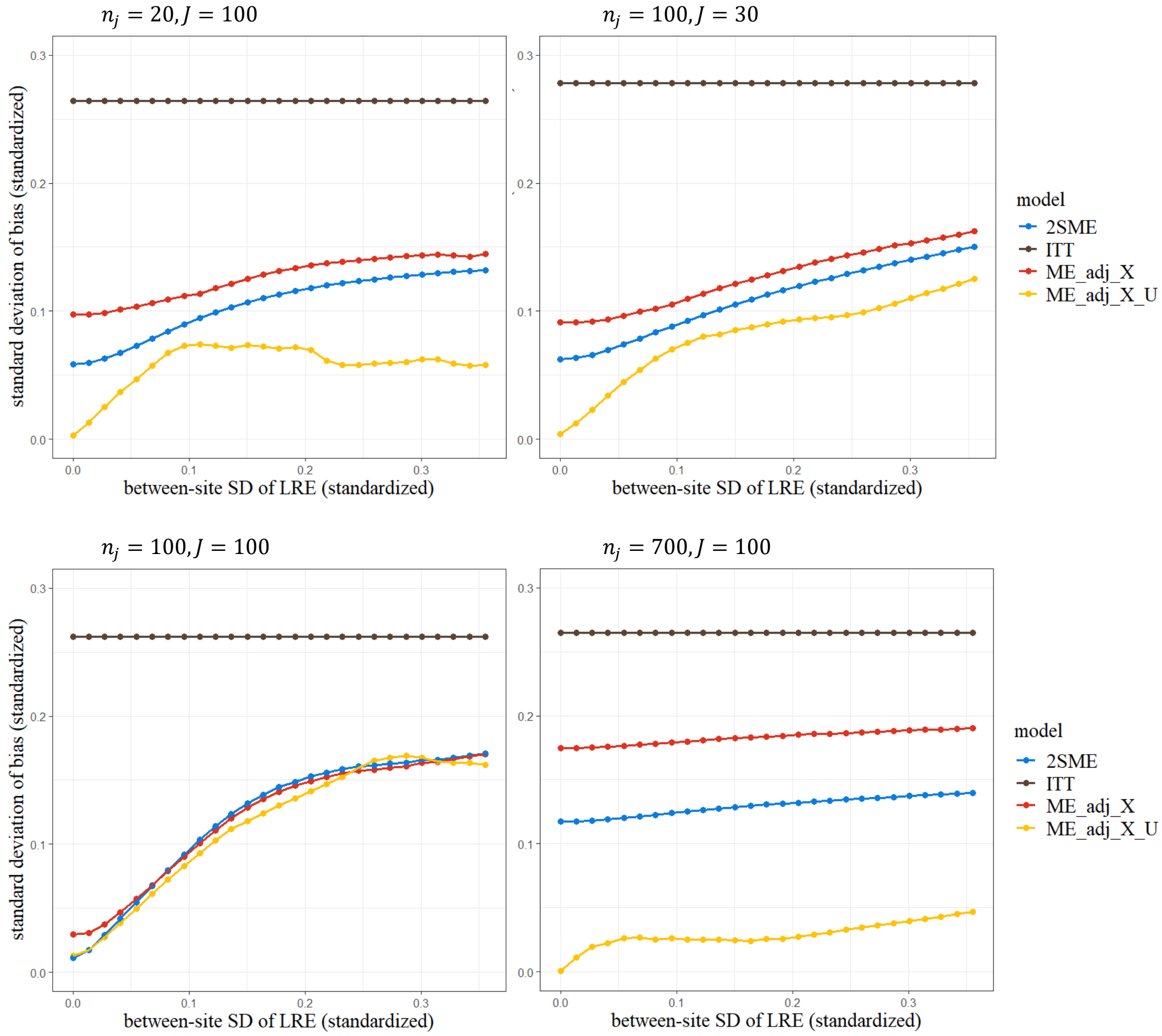

Figure S8. Between-Model Comparison of the Standard Deviation of Bias in LRE Estimation Under Scenario 2

## ***Efficiency***

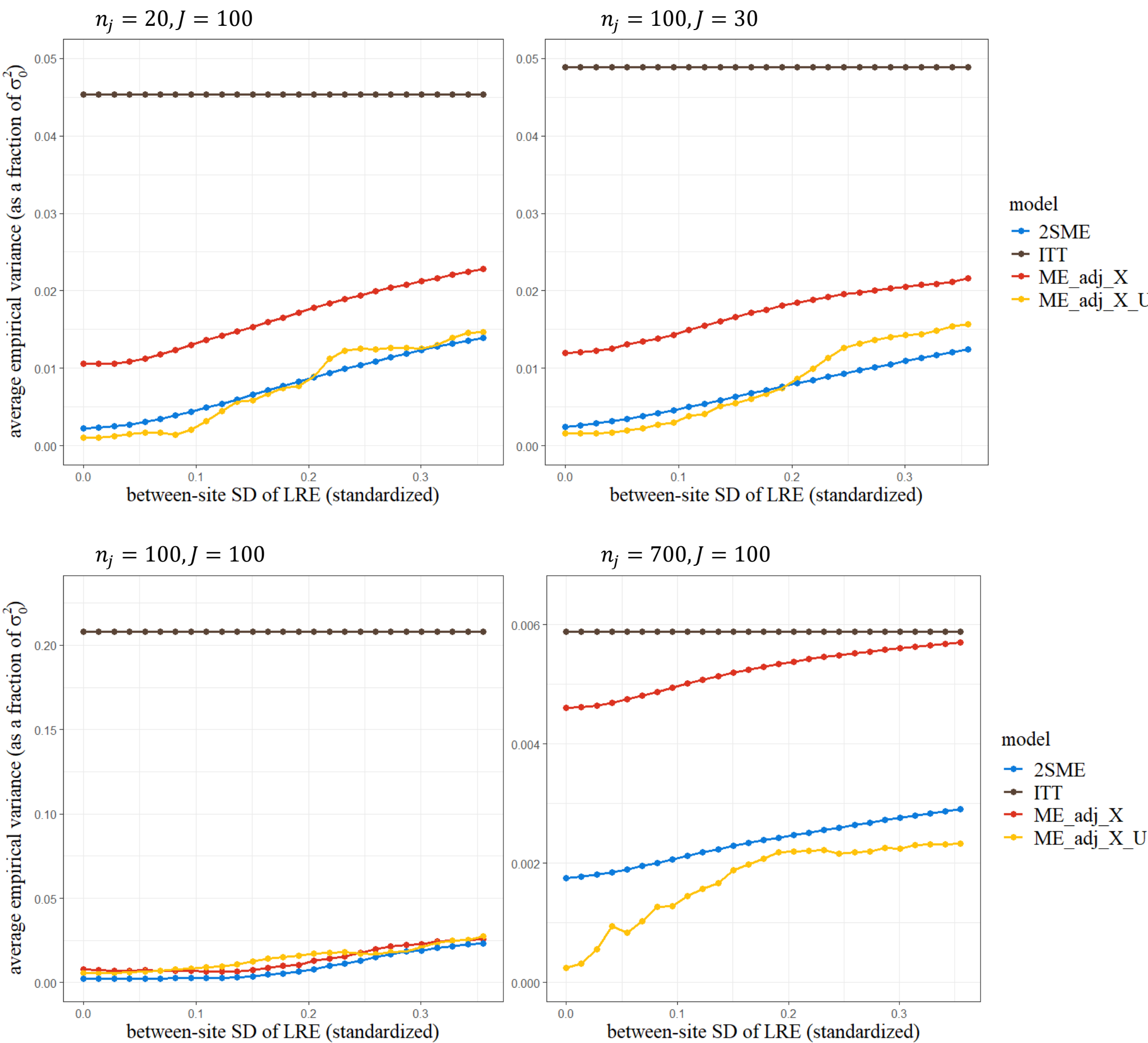

Note: The vertical scales differ between the three sample sizes per site.

Figure S9. Between-Model Comparison of Efficiency in LRE Estimation Under Scenario 2

## ***RMSE***

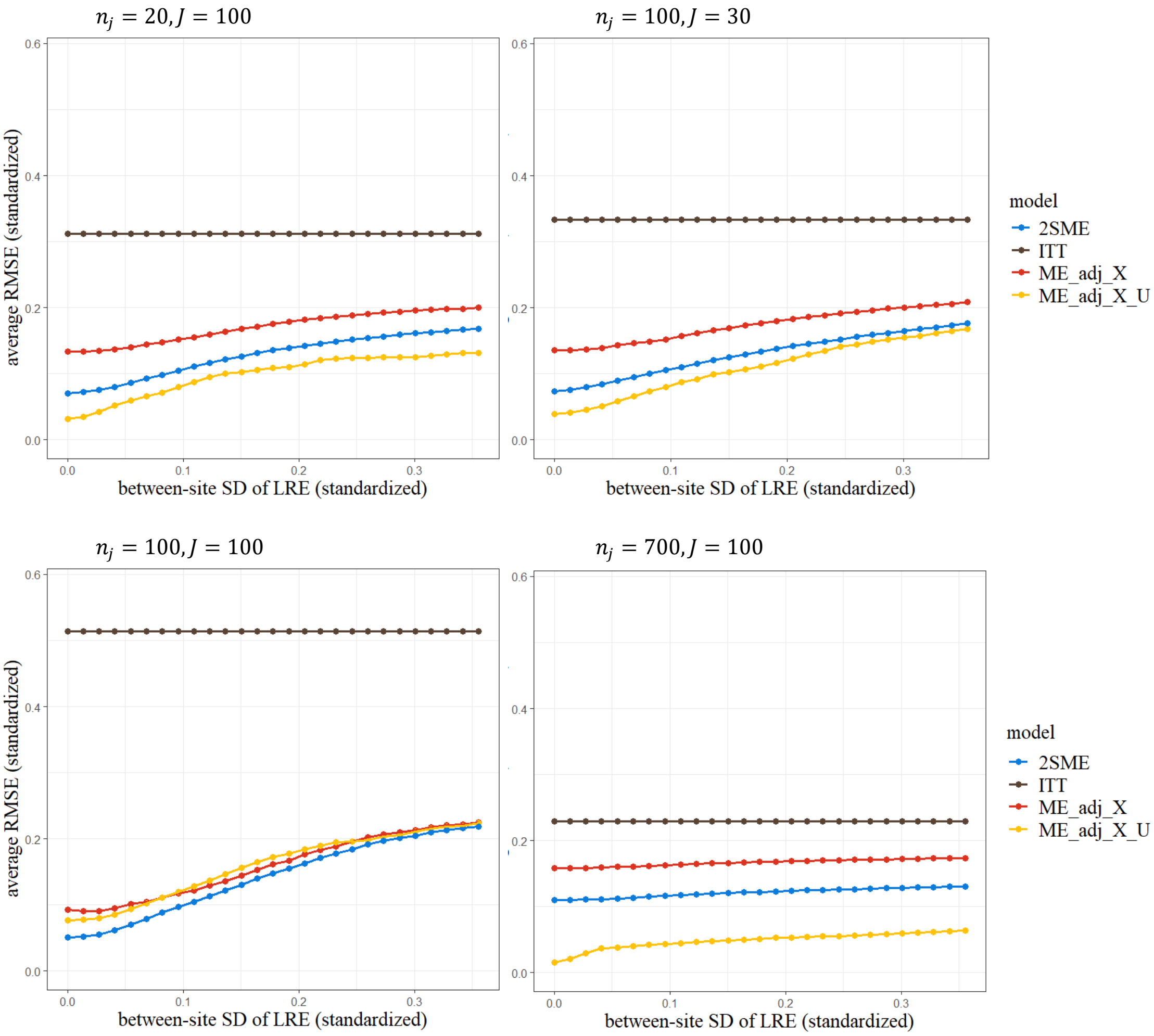

Figure S10. Between-Model Comparison of RMSE in LRE Estimation Under Scenario 2

## ***SCE Rate***

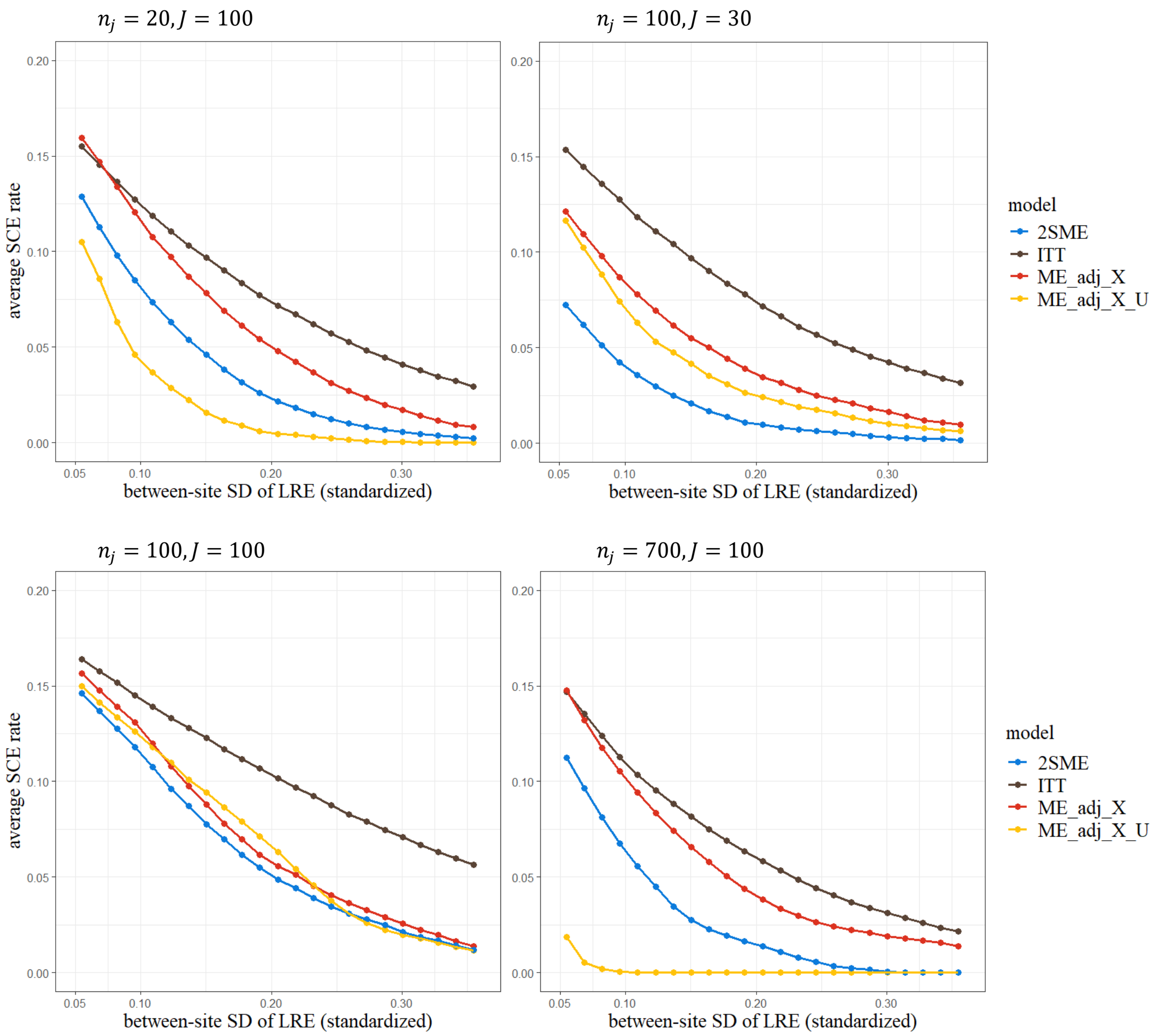

Figure S11. Between-Model Comparison of the SCE Rate in LRE Estimation Under Scenario 2

### *MCE Rate*

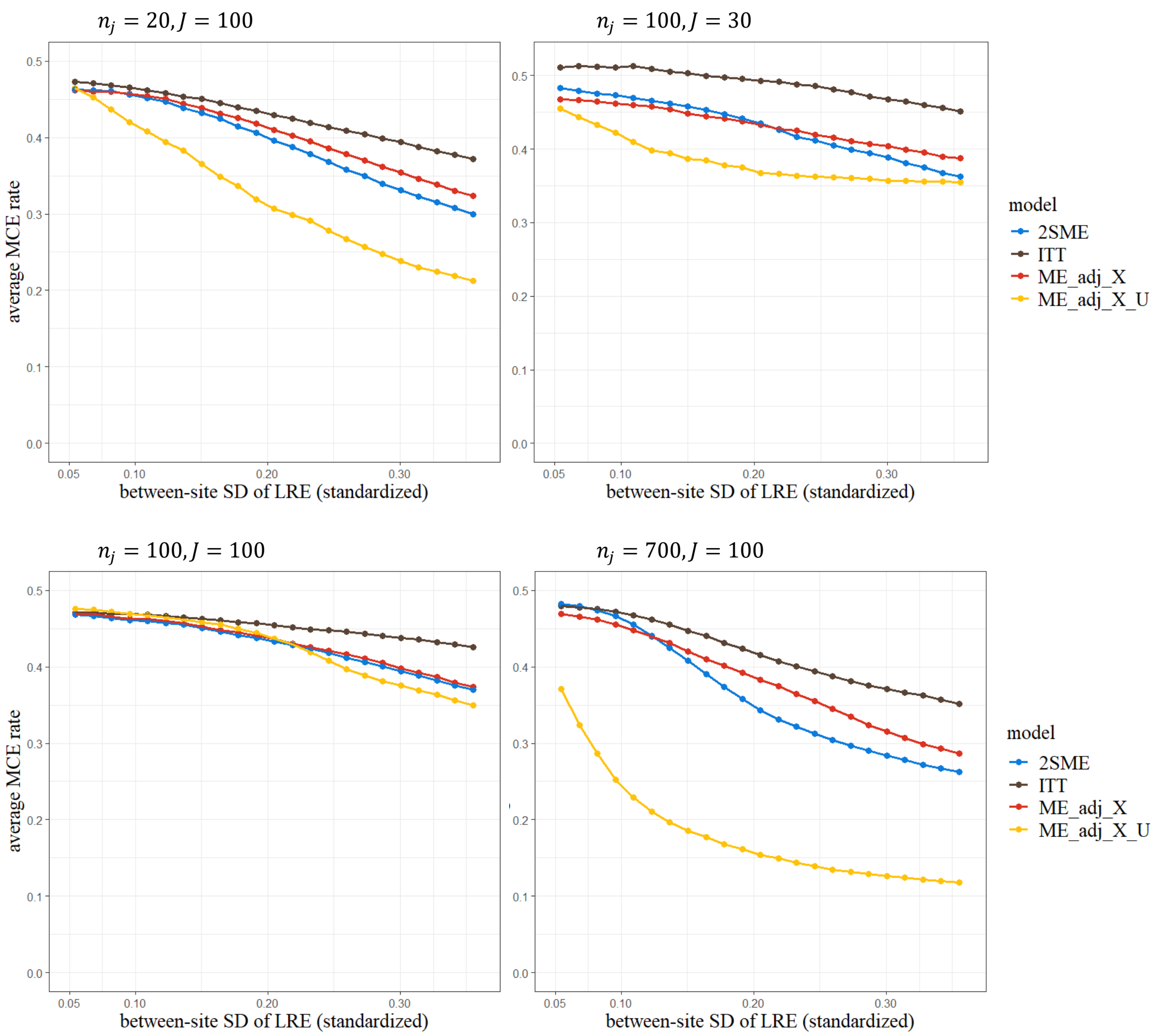

Figure S12. Between-Model Comparison of the MCE Rate in LRE Estimation Under Scenario 2

## II.d. Testing the Consistency of the 2SME Estimator

To empirically examine consistency as a theoretical property of the 2SME estimator, we compute the average absolute bias and the standard deviation of bias over the sampled sites for scenario 1 when the violation of Assumption 1 is minor. Due to computational constraints, these results are based on one simulated data file. We have made some small modifications to the data generation process: To reduce unnecessary noise, we let the distribution of the error term for $Y(0)$ be $N(0,1)$. Thus, the within-site SD of $Y(0)$ is approximately 175. We also reduce the within-site variance of $\delta_{ij}$, changing it from 4 to 1, thereby reducing the error variance in $Y(1)$.

The simulation results are summarized in Table S1 below. The R package "lme4" for mixed-effects modeling, however, cannot handle a data structure with 10,000 sites and 10,000 or 50,000 individuals per site. The results show the general trend that the bias decreases and becomes close to zero only when both the number of sites $J$ and the average sample size per site $n_j$ go to infinity. When either $J$ or $n_j$ is fixed at a relatively low level, increasing the magnitude of the other does not lead to a convergence to the true parameter values of LRE.

Table S1. Magnitude of Bias as a Function of the Number of Sites and the Sample Size per Site

| Average Absolute Bias | | | | Standard Deviation of Bias | | | |
|---|---|---|---|---|---|---|---|
| $\bar{n}_j$ | $J$ = 100 | $J$ = 1,000 | $J$ = 10,000 | $\bar{n}_j$ | $J$ = 100 | $J$ = 1,000 | $J$ = 10,000 |
| 100 | 0.075 | 0.073 | 0.076 | 100 | 0.097 | 0.091 | 0.095 |
| 1,000 | 0.037 | 0.031 | 0.030 | 1,000 | 0.045 | 0.039 | 0.038 |
| 10,000 | 0.027 | 0.021 | × | 10,000 | 0.035 | 0.026 | × |
| 50,000 | 0.026 | 0.019 | × | 50,000 | 0.034 | 0.024 | × |

Note: The standardized $\sqrt{var(\theta_j)} = 0.1$.

| Average Absolute Bias | | | | Standard Deviation of Bias | | | |
|---|---|---|---|---|---|---|---|
| $\bar{n}_j$ | $J$ = 100 | $J$ = 1,000 | $J$ = 10,000 | $\bar{n}_j$ | $J$ = 100 | $J$ = 1,000 | $J$ = 10,000 |
| 100 | 0.106 | 0.084 | 0.083 | 100 | 0.140 | 0.105 | 0.104 |
| 1,000 | 0.078 | 0.038 | 0.032 | 1,000 | 0.098 | 0.047 | 0.041 |
| 10,000 | 0.068 | 0.026 | × | 10,000 | 0.089 | 0.032 | × |
| 50,000 | 0.069 | 0.025 | × | 50,000 | 0.091 | 0.031 | × |

Note: The standardized $\sqrt{var(\theta_j)} = 0.35$.

# III. Supplementary Process and Results of NJCS Data Analysis

## III.a. Covariate Selection and Model Specification

Firstly, when all the theoretically important covariates in the first set are controlled for, we find that adopting a cross-validation (CV) LASSO procedure or applying a relatively stringent "one-standard-error" rule (1 SE) to the procedure does not lead to a notable difference in the LRE estimates. Secondly, the models that adjust for the first set of covariates may include either the site-level aggregates of categorical measures of age, baseline earnings, and employment to allow for nonlinearity or the aggregated continuous measures of these covariates assuming linearity. However, the specifications with the continuous covariate measures show multicollinearity with severely inflated coefficient estimates and standard errors. Thirdly, for the primary specification that includes both the EB estimate of the control group mean outcome and the other site-level baseline covariates, we notice a strong association between the control group average earnings in year 4 and the control group employment rate in year 2. In fact, the specifications that include or exclude the control group employment rate in year 2 lead to identical LRE estimates. These results suggest that the labor market prospects at each site reflected in the control group employment rate in year 2 is fully captured by the control group average earnings in year 4. Therefore, our primary specification adjusts for the first-set covariates including the categorical measures of age, baseline earnings, and employment while excluding the control group employment rate in year 2. We employ the stringent 1 SE LASSO procedure for covariate selection from the second set to keep the model relatively parsimonious.

## III.b. Additional NCJS Analytic Results

Figure S13 displays empirical bayes (EB) estimates of the site-specific ITT effects for the 100 Job Corps centers, obtained from a random-intercept random-slope mixed-effects model with no covariates.

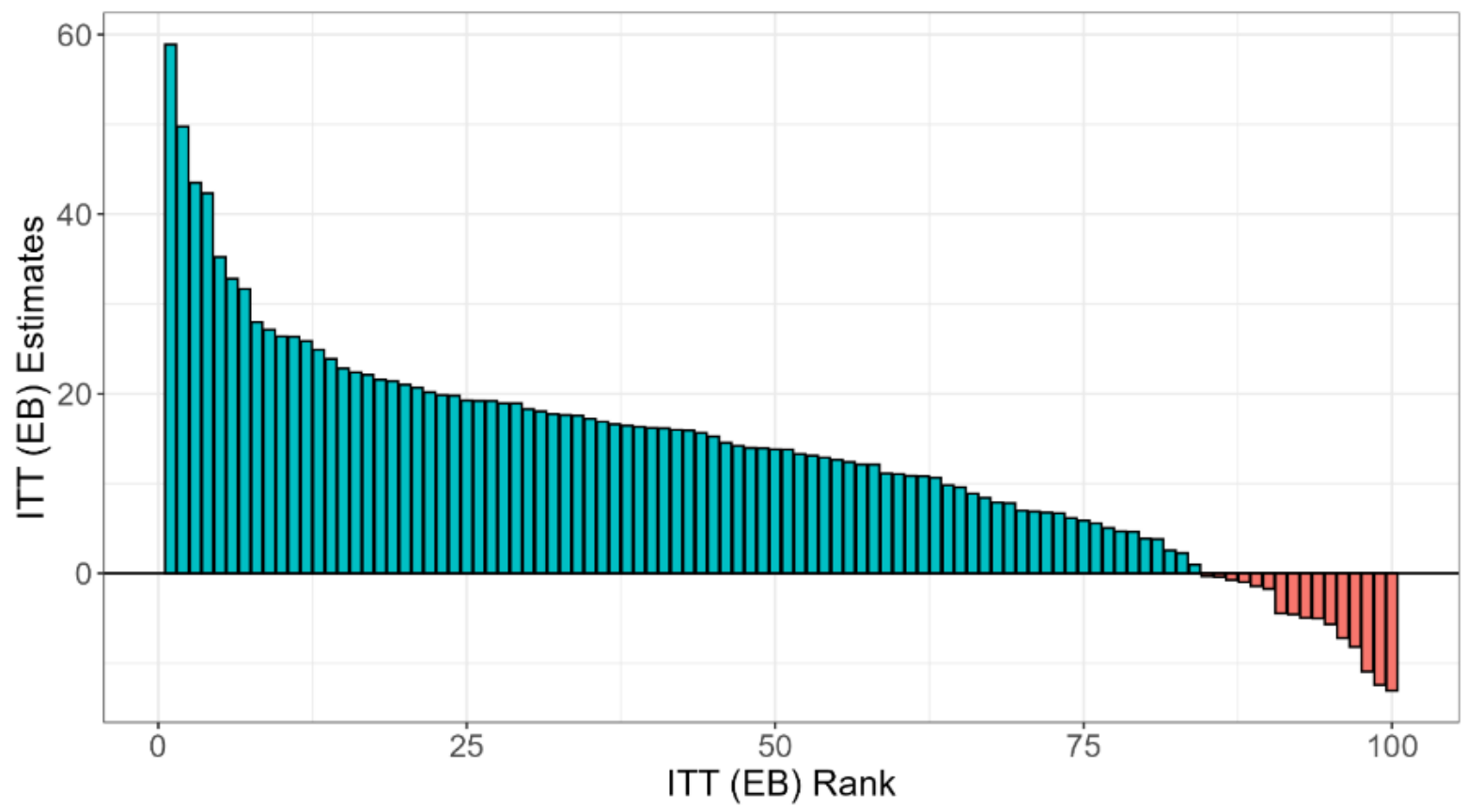

*Note*: Red bars indicate Job Corps centers with site-specific ITT effect estimates below zero.

Figure S13. Empirical Bayes Estimates of the Unadjusted Site-Specific ITT Effects of the One Hundred Job Corps Centers.

Figure S14 shows that Job Corps center rankings based on their estimated ITT effects are misaligned with their LRE rankings.

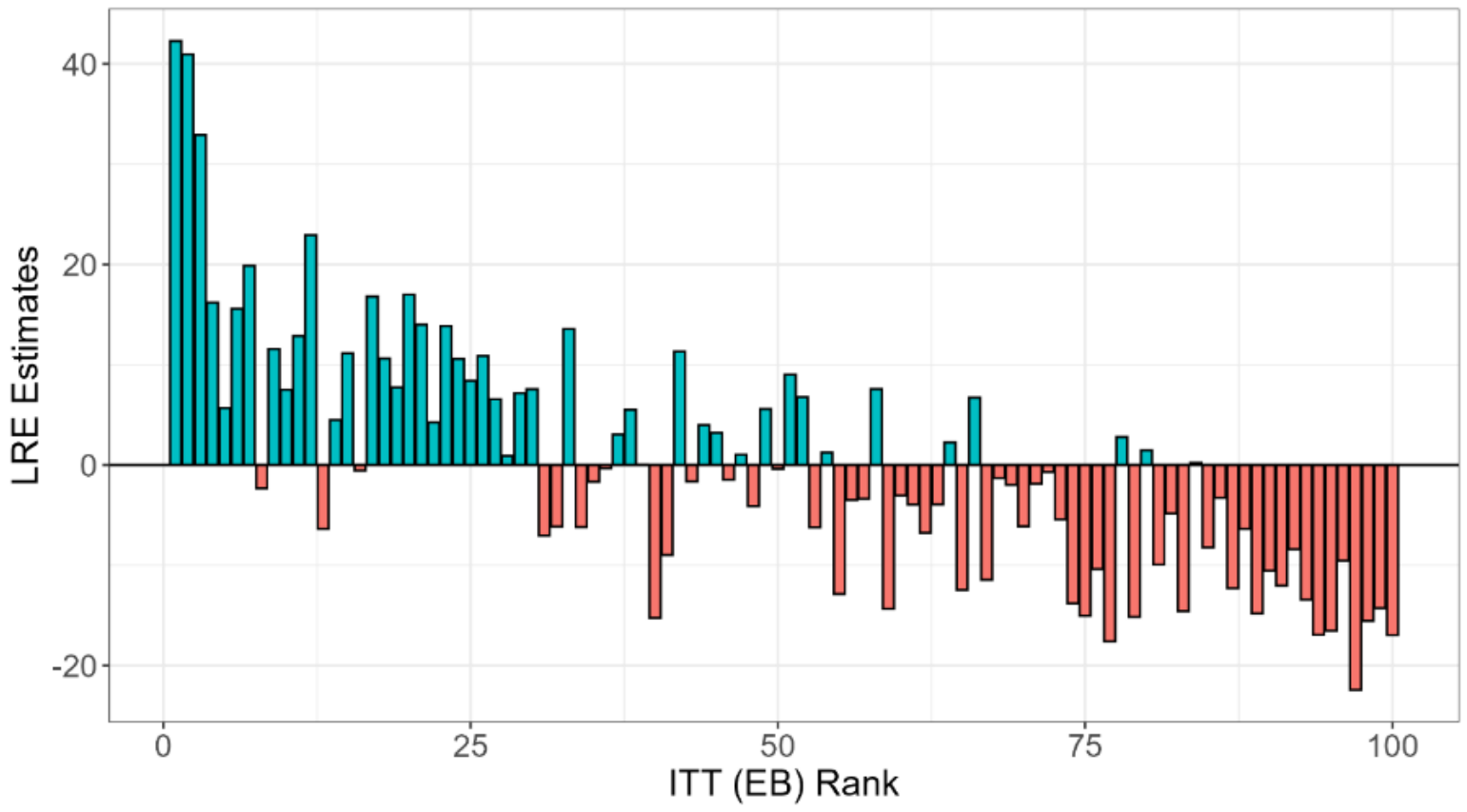

*Note*: Red bars indicate Job Corps centers with LRE estimates below zero.

Figure S14. Estimated LRE of Job Corps Centers Ordered by Their Rankings Based on the Empirical Bayes Estimates of the Unadjusted Site-Specific ITT Effects.